\documentclass[draft]{agujournal2019}
\usepackage{url} 
\usepackage{lineno}
\usepackage[finalnew]{trackchanges}
\usepackage{soul}
\usepackage{romannum}
\usepackage{amsmath}
\usepackage{amsfonts}
\usepackage{amssymb}

\draftfalse

\journalname{JGR: Space Physics}

\begin{document}

%
%


\title{Nonlinear Landau resonant interaction between kinetic Alfv\'en waves and thermal electrons: Excitation of time domain structures}

%
%




\authors{Xin An\affil{1}, Jacob Bortnik\affil{1}, Xiaojia Zhang\affil{2, 1}}


\affiliation{1}{Department of Atmospheric and Oceanic Sciences, University of California, Los Angeles, CA, USA}
\affiliation{2}{Department of Earth, Planetary and Space Sciences, University of California, Los Angeles, CA, USA}




\correspondingauthor{Xin An}{xinan@atmos.ucla.edu}




\begin{keypoints}
\item Electrons in nonlinear Landau resonance with the kinetic Alfven wave form a spatially modulated beam distribution and excite TDSs.
\item The phase mixing rate dominates over TDS growth rate at a large enough wave potential and an upper bound of wave potential is obtained.
\item A big picture emerges of energy cascading from dipolarization fronts to TDSs ending up in the thermal electrons.
\end{keypoints}

%
%

%
%


\begin{abstract}
Phase space holes, double layers and other solitary electric field structures, referred to as time domain structures (TDSs), often occur around dipolarization fronts in the Earth's inner magnetosphere. They are considered to be important because of their role in the dissipation of the injection energy and their potential for significant particle scattering and acceleration. Kinetic Alfv\'en waves are observed to be excited during energetic particle injections, and are typically present in conjunction with TDS observations. Despite the availability of a large number of spacecraft observations, the origin of TDSs and their relation to kinetic Alfv\'en waves remains poorly understood to date. Part of the difficulty arises from the vast scale separations between kinetic Alfv\'en waves and TDSs. Here, we demonstrate that TDSs can be excited by electrons in nonlinear Landau resonance with kinetic Alfv\'en waves. These electrons get trapped by the parallel electric field of kinetic Alfv\'en waves, form localized beam distributions, and subsequently generate TDSs through beam instabilities. A big picture emerges as follows: macroscale dipolarization fronts first transfer the ion flow (kinetic) energy to kinetic Alfv\'en waves at intermediate scale, which further channel the energy to TDSs at the microscale and eventually deposit the energy to the thermal electrons in the form of heating. In this way, the ion flow energy associated with dipolarization fronts is effectively dissipated in a cascade from large to small scales in the inner magnetosphere.
\end{abstract}


%
%

%


%
%
%
%

\section{Introduction}
The term ``time domain structures'' (TDSs) refers to packets of $\geqslant 1$\,ms duration intense electric field spikes detected by Van Allen Probes in the Earth's outer radiation belt \cite{mozer2015time}. The spatial scale of TDSs (in the direction along the magnetic field lines) is on the order of tens of Debye Lengths ($\sim 0.3$ -- $0.6$\,km) \cite{vasko2017electron, malaspina2018census}. TDSs propagate at a velocity comparable to the electron thermal velocity and are usually identified as electron acoustic-like mode \cite{vasko2017electron2}. TDSs have significant electric field components parallel to the local background magnetic field. Depending on the appearance of these parallel electric fields, TDSs are generally categorized into double layers and phase space holes. The double layer has a unipolar parallel electric field \cite<e.g.,>[]{quon1976formation}, which resembles a net potential drop from a two-layer structure composed of a layer of net positive charges to an adjacent layer of net negative charges; The phase space hole has a bipolar electric field, which resembles the field created by a collection of positive or negative charges \cite<e.g.,>[]{schamel1979theory}. TDSs can interact with thermal electrons in the energy range between tens of eV to a few keV, causing efficient pitch angle scattering and acceleration of thermal electrons \cite{osmane2014threshold, artemyev2014thermal, vasko2017diffusive}.

Statistical observations by the Van Allen Probes have shown that TDSs permeate the inner magnetosphere \cite{malaspina2014nonlinear}. The occurrence of these TDSs in the inner magnetosphere is strongly correlated with macroscopic plasma boundaries, such as dipolarization fronts \cite{malaspina2015electric}. Also, the regions where TDSs are excited travel with propagating dipolarization fronts. These observations indicate that the free energy of the plasma boundary motion at the macroscale is transferred to that of TDSs at the microscale. But how does this energy transfer manifest itself between different spatial scales? The answer is related to whistler and kinetic Alfv\'en waves that have spatial scales between dipolarization fronts and TDSs. In fact, both of these waves are observed to be excited during energetic particle injections into the inner magnetosphere \cite{chaston2014observations, malaspina2018census}, and are typically present in TDS observations \cite{malaspina2015electric, mozer2015time, chaston2015broadband}.

It was suggested that the localized electric field associated with kinetic Alfv\'en waves may locally accelerate electrons to form an electron beam \cite<e.g.,>[]{damiano2015ion, artemyev2015electron, damiano2016ion}, leading to TDS excitation through beam instabilities \cite<e.g.,>{malaspina2015electric}. The same hypothesis has been proposed to explain the abundance of double layers and phase space holes in the bursty bulk flow braking region of the magnetotail \cite<e.g.,>{stawarz2015generation, ergun2015large, chaston2012energy}. \add{In the auroral ionosphere, despite it being a low beta plasma, the generation of double layers and phase space holes has been attributed to beam instabilities driven by the parallel electric field of inertial Alfv\'en waves} \cite{silberstein1994computer, genot2004alfven}. In connection with this hypothesis, the parallel electric field of whistler waves was observed to be strong enough to accelerate Landau resonant electrons and generate Langmuir waves \cite{reinleitner1982chorus, li2017chorus}. It was further demonstrated that Landau resonant electrons trapped by the whistler parallel electric field generate a range of TDSs, and that a single quantity, the ratio of Landau resonant velocity to electron thermal velocity, controls the type of TDSs that will be generated \cite{an2019unified}. \add{It is worth mentioning that an alternative scenario was proposed for the formation of electric field spikes, such as nonlinear fluid steepening of electron acoustic modes} \cite{vasko2018electrostatic, agapitov2018nonlinear}. These studies motivate us to further investigate TDSs driven by kinetic Alfv\'en waves. Concretely, we aim to demonstrate that TDSs can be excited by Landau resonant electrons trapped in kinetic Alfv\'en waves through beam instabilities. In Section \ref{sec-linear-kinetic}, we will analyze properties of kinetic Alfv\'en waves using linear kinetic theory, which then serve as the initialization for particle-in-cell simulations. In Section \ref{sec-tds-excitation}, using particle-in-cell simulations, we will demonstrate how kinetic Alfv\'en waves drive TDSs by electron phase trapping and show the associated characteristics of TDSs. With the intuition gained from particle-in-cell simulations, we will also analyze the critical condition for TDS excitation. In Section \ref{sec-conclusions}, we summarize our results and put our work in the bigger context regarding energy dissipation around dipolarization fronts.

\section{Linear dispersion relation of kinetic Alfv\'en waves}\label{sec-linear-kinetic}
To familiarize ourselves with the linear properties of kinetic Alfv\'en waves and prepare for the initialization of particle-in-cell simulations, we start with the linear kinetic theory. The linearized Vlasov equation is combined with Maxwell's equations to solve for the hot plasma dispersion relation. The solution is found via the following equation, in matrix form,
\begin{linenomath*}
	\begin{equation}\label{eq-dispersion-matrix-form}
		\begin{pmatrix}
		  -n_z^2 + \epsilon_{xx}         &                  \epsilon_{xy}           &     n_x n_z + \epsilon_{xz}  \\
		           \epsilon_{yx}         & -n_x^2 - n_z^2 + \epsilon_{yy}           &               \epsilon_{yz}  \\
	     n_x n_z + \epsilon_{zx}         &                  \epsilon_{zy}           &      -n_x^2 + \epsilon_{zz}
		\end{pmatrix}
		\begin{pmatrix}
		\delta E_x \\
		\delta E_y \\
		\delta E_z
		\end{pmatrix}  = \mathbf{0} ,
	\end{equation}
\end{linenomath*}
where $\mathbf{n} = (n_x, n_y, n_z)$ is the refractive index and the coordinate system in which $n_y = 0$ is chosen. The background magnetic field is in the $z$ direction. All the complexities of evaluating velocity space integrals are contained in elements of the dielectric tensor $\epsilon_{ij}$, the derivation of which is fairly standard and can be found in textbooks \cite<e.g.,>{stix1992waves, swanson2012plasma, ichimaru2018basic}. For any nontrivial solution of the wave field $(\delta E_x, \delta E_y, \delta E_z)$, the determinant of dispersion matrix is required to be zero. To this end, we use the hot plasma dispersion relation solver in the HOTRAY code \cite{horne1989path} to find the roots of the determinant.

We consider a collisionless, homogeneous plasma, which is composed of electrons and protons for simplicity. The proton-to-electron mass ratio is $m_i / m_e = 1836$. The normalized Alfv\'en velocity is $v_A / c = \omega_{ci} / \omega_{pi} = \frac{1}{5} \times \sqrt{\frac{1}{1836}} = 4.67 \times 10^{-3}$. Here $c$ is the speed of light, $\omega_{ci}$ is the proton cyclotron frequency, and $\omega_{pi}$ is the proton plasma frequency. Each species is assumed to have a Maxwellian velocity distribution\add{, which is a simplification from realistic situations} \cite{vasko2017electron2, walsh2020census}. The proton and electron thermal velocities are $v_{Ti} = v_A / 8$ and $v_{Te} = 4 v_A$, respectively. \add{These values are chosen to represent typical conditions around dipolarization fronts, i.e., electron temperature $T_e = 300$\,eV, ion temperature $T_i = 2000$\,eV, background magnetic field $B_0 = 100$\,nT, plasma density $n_0 = 1\,\text{cm}^{-3}$.} \remove{Here the ion and electron thermal velocities are representative of conditions around dipolarization fronts penetrating deep into the equatorial region of the Earth's inner magnetosphere, where the cold plasma population significantly exceeds the hot, injected plasma population.} We focus on the wave dispersion properties and thus did not include any free energy source for wave excitation. We note that, in the limit $v_{Te} > v_A$, the shear Alfv\'en wave is termed the ``kinetic Alfv\'en wave'' \cite<e.g.,>{hasegawa1976particle, lysak1996kinetic}. In this kinetic regime, the characteristic perpendicular length scale is the ion acoustic gyroradius $\rho_s = c_s / \omega_{ci}$, where $c_s$ is the ion acoustic speed. In the opposite limit $v_{Te} < v_A$, the electron inertia becomes important and the shear Alfv\'en wave is termed the ``inertial Alfv\'en wave'' \cite<e.g.,>{goertz1979magnetosphere}. The inertial regime is applicable to the Earth's ionosphere. In this study, however, we focus on the kinetic regime which is representative of dipolarization fronts located in the equatorial magnetosphere. It is worthy to clarify that the nomenclature ``electromagnetic ion cyclotron waves'' (EMIC) is often used to refer to shear Alfv\'en waves when the wave frequency is relatively large and does not satisfy the condition $\omega \ll \omega_{ci}$, regardless of the kinetic or inertial regime. Here we use ``kinetic Alfv\'en waves'' to unambiguously refer to shear Alfv\'en waves in the limit $v_{Te} > v_A$, regardless of the wave frequency.

\begin{figure}[tphb]
	\centering
	\includegraphics[width=4.5in]{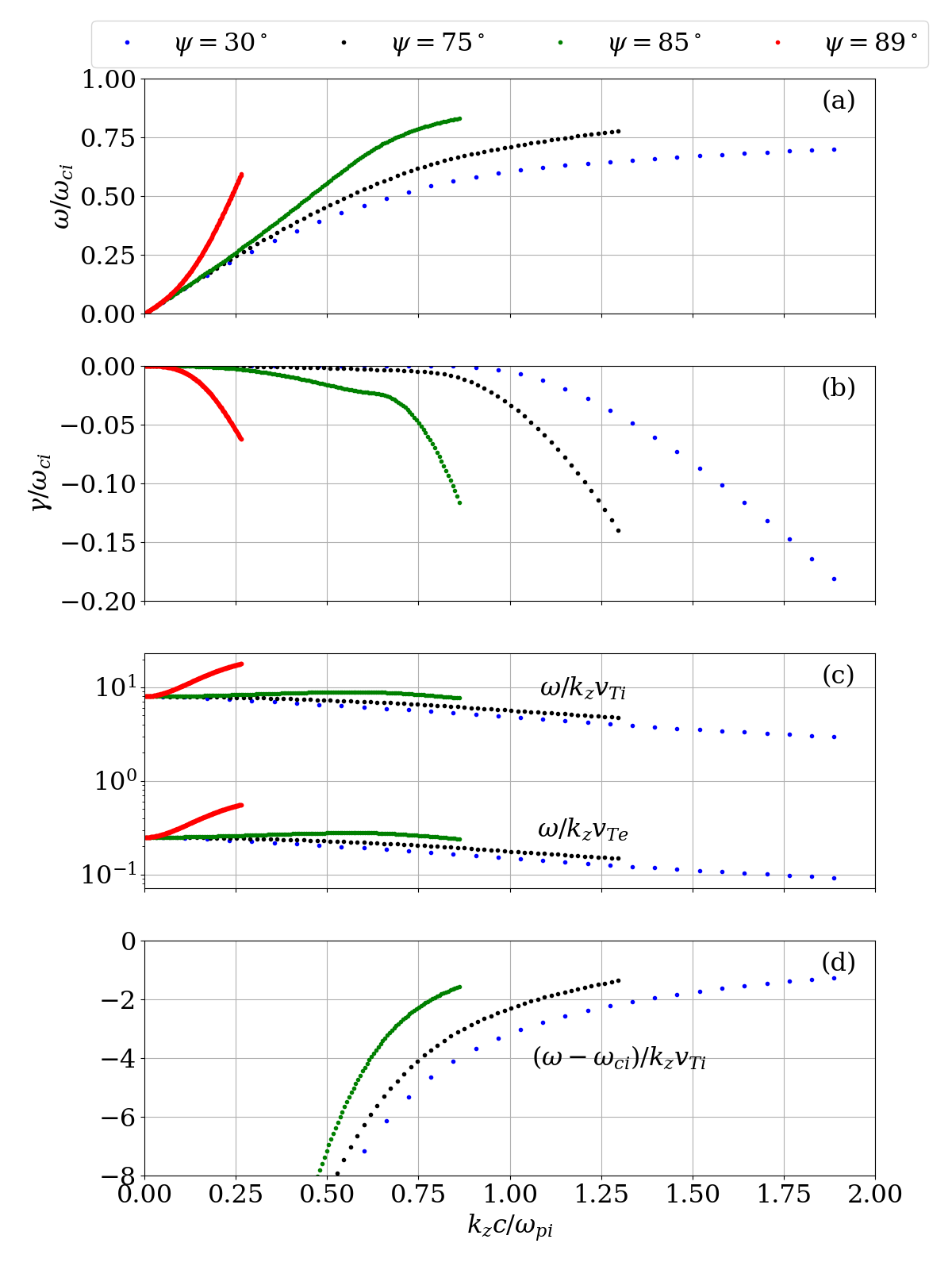}
	\caption{The dispersion relation of kinetic Alfv\'en waves for four representative wave normal angles. In each panel, $\psi=30^{\circ},75^{\circ},85^{\circ},89^{\circ}$ correspond to blue, black, green and red dots, respectively. All physical quantities on the vertical axis are shown as a function of normalized parallel wavenumber $k_z c / \omega_{pi}$ on the horizontal axis. (a) Normalized wave frequency $\omega / \omega_{ci}$. (b) Normalized wave growth rate $\gamma / \omega_{ci}$. Note that $\gamma<0$ indicates wave damping. (c) Parallel phase velocity or Landau resonant velocity $\omega / k_z$ compared to ion and electron thermal velocities. (d) Ion cyclotron resonant velocity. Note that ions in cyclotron resonance counter-propagate with the wave ($k_z > 0$) and thus have negative cyclotron resonant velocities. The ion cyclotron resonant velocity for $\psi = 89^\circ$ is out of range.}
	\label{fig-linear-kinetic-dr}
\end{figure}

The hot plasma dispersion relation of kinetic Alfv\'en waves ranging from slightly oblique ($\psi = 30^\circ$) to almost perpendicular ($\psi = 89^\circ$) propagation are shown in Figure \ref{fig-linear-kinetic-dr}. Here $\psi$ stands for the wave normal angle between the wavenumber vector $\mathbf{k} = (k_x, 0, k_z)$ and the background magnetic field $\mathbf{B}_0 = (0, 0, B_0)$. In the long wavelength limit $|\mathbf{k}| \to 0$, shear Alfv\'en waves at different propagation angles converge to $\omega = k_z v_A$ [Figure \ref{fig-linear-kinetic-dr}(a)]. The damping rate is negligibly small in this limit [Figure \ref{fig-linear-kinetic-dr}(b)]. As wavenumber increases, the parallel phase velocity $\omega / k_z$ decreases for slightly oblique propagation ($\psi = 30^\circ$), whereas $\omega / k_z$ increases for highly oblique propagation ($\psi = 89^\circ$). This distinct characteristic of $\omega / k_z$ at different wave normal angles has important implications for the wave damping mechanism \cite<e.g.,>{gary2004kinetic}. In the latter case of highly oblique propagation (e.g., $\psi = 89^\circ$), where $\omega / k_z$ approaches the electron thermal velocity with increasing wavenumber [Figure \ref{fig-linear-kinetic-dr}(c)], electron Landau resonance is responsible for the damping of kinetic Alfv\'en waves in the small wavelength limit [Figures \ref{fig-linear-kinetic-dr}(b) and \ref{fig-linear-kinetic-dr}(c)]. In the former cases (e.g., $\psi = 30^\circ, 75^\circ, 85^\circ$), where $\omega / k_z$ stays away from $v_{Te}$ with increasing wavenumber [Figure \ref{fig-linear-kinetic-dr}(c)] while the ion cyclotron resonant velocity approaches the ion thermal velocity [Figure \ref{fig-linear-kinetic-dr}(d)], ion cyclotron damping becomes more effective in the small wavelength limit [Figures \ref{fig-linear-kinetic-dr}(b) and \ref{fig-linear-kinetic-dr}(d)]. In all the present cases, the contribution of ion Landau damping is relatively small, since $\omega / k_z$ is large compared to ion thermal velocity [i.e., $v_{Ti} = v_A / 8$; see Figure \ref{fig-linear-kinetic-dr}(c)]. We can imagine that, if the ion thermal velocity approaches the Alfv\'en velocity, ion Landau damping may become more pronounced.

\begin{figure}[tphb]
	\centering
	\includegraphics[width=4.5in]{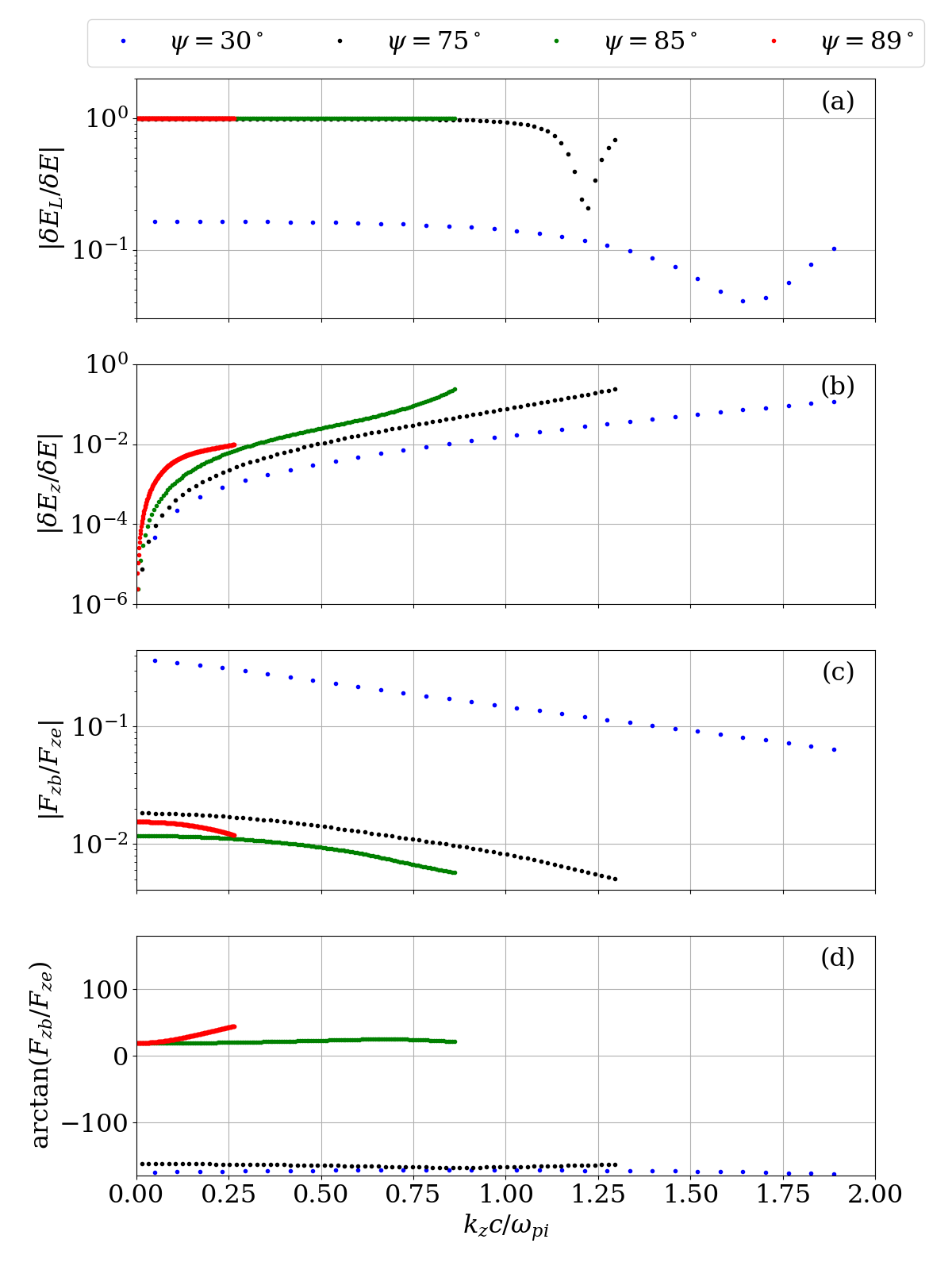}
	\caption{Properties of kinetic Alfv\'en waves derived from the linear kinetic theory. The format of this figure is the same as that of Figure \ref{fig-linear-kinetic-dr}. (a) The ratio of electrostatic field $\delta E_L$ to total electric field $\delta E$. (b) The ratio of parallel electric field $\delta E_z$ to total electric field $\delta E$. (c, d) The magnitude and phase of the ratio of the wave magnetic mirror force $F_{zb}$ to the parallel electric force $F_{ze}$.}
	\label{fig-linear-kinetic-properties}
\end{figure}

After solving for the hot plasma dispersion relation, we can further obtain ratios between different components of electric fields [Figure \ref{fig-linear-kinetic-properties}]. The ratio of wave electrostatic field $\delta E_L$ to total electric field $\delta E$ is shown in Figure \ref{fig-linear-kinetic-properties}(a). Here $\delta E_L$ is the projection of wave electric field onto the $\mathbf{k}$ direction. In the limit of parallel propagation ($\psi = 0$), the wave is purely electromagnetic, i.e., $\delta E_L = 0$ (not shown). At slightly oblique propagation ($\psi = 30^\circ$), a small portion ($|\delta E_L / \delta E| \approx 0.15$) of the total electric field is electrostatic, whereas the waves are nearly electrostatic ($|\delta E_L / \delta E| \approx 1$) at $\psi = 75^{\circ}, 85^{\circ}, 89^{\circ}$. The ratio of parallel electric field $\delta E_z$ to total electric field $\delta E$ is shown in Figure \ref{fig-linear-kinetic-properties}(b). It is seen that the parallel electric field is only a very small portion of the total electric field at all propagation angles, i.e., $|\delta E_z / \delta E| < 10^{-2}$ for wave modes that are not heavily damped.

The parallel electric field of kinetic Alfv\'en wave can trap electrons in its potential well due to its finite amplitude, which is the so-called nonlinear Landau resonance \cite<e.g.,>{o1965collisionless}. Even though the magnitude of $\delta E_z$ is small for kinetic Alfv\'en waves, the integral of this electric field over an extended wavelength along the background magnetic field can still be significant \cite{artemyev2015electron, artemyev2017nonlinear}. Indeed, the half width of trapping island can be compared to the electron thermal velocity as
\begin{linenomath*}
	\begin{equation}\label{eq-trapping-island-width}
		\frac{\Delta v_{tr}}{v_{Te}} = \frac{2 \omega_{tr} / k_z}{v_{Te}} = \frac{2}{k_z v_{Te}} \sqrt{\frac{e k_z \cdot k_z \delta\phi}{m_e}} = 2 \sqrt{\frac{e \delta \phi}{T_e}} ,
	\end{equation}
\end{linenomath*}
where $\omega_{tr} = \sqrt{e k_z |\delta E_z| / m_e}$ is the oscillation frequency for electrons trapped at the bottom of the potential well (i.e., the trapping frequency), and $T_e = m_e v_{Te}^2$ is the electron temperature. The wave potential $\delta \phi$ is defined through the relation $\delta E_z = -i k_z \delta \phi$, where $i$ is the unit imaginary number. Evidently, it is the ratio of the magnitude of wave potential $e \delta \phi$ to the electron temperature $T_e$ that controls the significance of electron trapping. \add{In addition, kinetic Alfv\'en waves around dipolarization fronts typically have a finite bandwidth $\Delta f$ in frequency or $\Delta \lambda$ in wavelength. For electron trapping to occur, the kinetic Alfv\'en waves should be coherent over the electron trapping time scale. This requires the wave coherence time $\tau_{\text{coherence}} = 1 / \Delta f = \Delta \lambda / (v_A \lambda^2)$ to be greater than the electron trapping time $\tau_{\text{trapping}} = \sqrt{m_e / (e k_z \vert\delta E_{z}\vert)}$, which needs to be verified by measurements.}

Apart from the parallel electric field, the parallel magnetic field perturbation $\delta B_z$ is present for oblique wave normal angles. The variation of $\delta B_z$ along the background magnetic field, i.e., $\frac{\partial}{\partial z}\delta B_z$, acts as a mini-magnetic mirror on electrons. Here the first adiabatic invariant of electrons (i.e., the electron magnetic moment) is conserved because electrons gyrate about the magnetic field much faster than the equivalent temporal variation of $\delta B_z$. The magnetic mirror force and the parallel electric force can be in phase or out of phase, which add up to be either constructive or destructive effects. For comparison, we write these two forces explicitly as
\begin{linenomath*}
	\begin{eqnarray}
		F_{ze} & = & - e \delta E_z , \\
		F_{zb} & = & - \mu_e \frac{\partial \delta B_z}{\partial z} = - i \mu_e k_z \delta B_z ,
	\end{eqnarray}
\end{linenomath*}
where $\mu_e = m_e v_\perp^2 / (2 B_0)$ is the electron magnetic moment. We can express $\delta B_z$ in terms of the electric field using Faraday's Law, i.e., $\delta B_z = (k_x c/ \omega) \delta E_y$. The ratio of $F_{zb}$ to $F_{ze}$ can be rewritten as
\begin{linenomath*}
	\begin{equation}
		\frac{F_{zb}}{F_{ze}} = i k_z k_x \cdot \frac{\mu_e c}{e \omega} \cdot \frac{\delta E_y}{\delta E_z} = \frac{i}{2} \hat{k}_z \hat{k}_x \cdot \left( \frac{v_{Te}}{v_A} \right)^2 \cdot \frac{\omega_{ci}^2}{\omega \omega_{ce}} \cdot \frac{\delta E_y}{\delta E_z} ,
	\end{equation}
\end{linenomath*}
where $\hat{k}_{x, z} = k_{x, z} c /\omega_{pi}$ is the normalized wavenumber, $\omega_{ce}$ is the electron cyclotron frequency, and $\mu_e$ is written as as $\mu_e = \frac{1}{2} m_e v_{Te}^2 / B_0$. Using $\delta E_y / \delta E_z$ derived from the hot plasma dispersion relation, the magnitude and phase of $F_{zb} / F_{ze}$ are shown as a function of $k_z$ in Figures \ref{fig-linear-kinetic-properties}(c) and \ref{fig-linear-kinetic-properties}(d), respectively. On the one hand, at $\psi = 30^\circ$, $F_{zb}$ is a fraction of $F_{ze}$ (i.e., $|F_{zb} / F_{ze}| \sim 0.1$), and they are out of phase by $180$ degrees. Thus the wave magnetic mirror force cancels part of the parallel electric force, causing a reduction of the trapping width in Equation \eqref{eq-trapping-island-width}. More importantly, the wave magnetic mirror force depends on the perpendicular velocity of electrons. This gives rise to an asynchronization of trapped electrons with different perpendicular velocities, and thus causes additional phase mixing. We denote this scenario as the regime of weak nonlinear Landau resonant interaction. On the other hand, at large wave normal angles ($\psi = 75^\circ, 85^\circ, 89^\circ$), $F_{zb}$ is negligible compared to $F_{ze}$ and hence $F_{ze}$ acts coherently on trapped electrons. We denote this scenario as the regime of strong nonlinear Landau resonant interaction.

It is often difficult to get accurate measurements of the low frequency electric fields due to their small amplitudes and low frequencies. Thus it is difficult to obtain the half trapping width $\Delta v_{tr}$ in Equation \eqref{eq-trapping-island-width} directly from electric field measurements. To have an estimation of $\Delta v_{tr}$, we replace $e k_z \delta \phi$ in Equation \eqref{eq-trapping-island-width} with $\lvert F_{ze} + F_{zb} \rvert = \mu_e k_z \delta B_z \cdot \lvert (F_{ze} + F_{zb}) / F_{zb} \rvert$, which takes the wave magnetic mirror force into account and gives
\begin{linenomath*}
	\begin{equation}\label{eq-vtr-practice}
		\frac{\Delta v_{tr}}{v_{Te}} = \sqrt{\frac{2 \delta B_z}{B_0} \cdot \bigg\lvert \frac{F_{ze} + F_{zb}}{F_{zb}} \bigg\rvert} .
	\end{equation}
\end{linenomath*}
In practice, we have accurate measurements of the wave magnetic fields and background plasma parameters, which give reasonable approximations of $\delta B_z / B_0$ and $F_{zb} / F_{ze}$ (from the linear kinetic theory). We will use Equation \eqref{eq-vtr-practice} to estimate $\Delta v_{tr}$ in our particle-in-cell simulations.

\section{TDS excitation by electron phase trapping}\label{sec-tds-excitation}
\subsection{Computational setup}
\begin{figure}[tphb]
	\centering
	\includegraphics[width=3in]{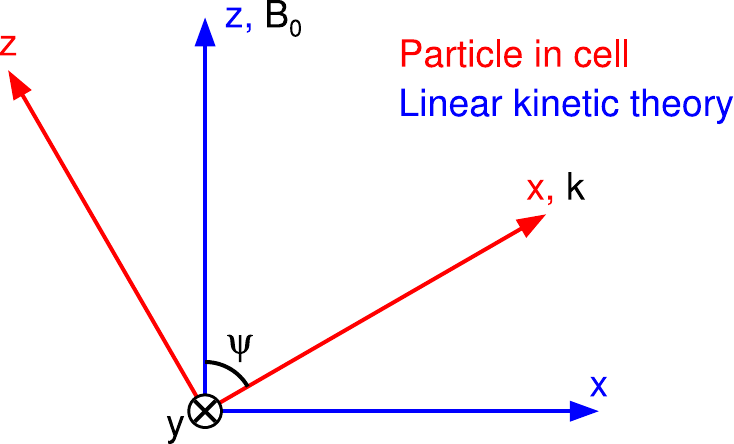}
	\caption{Comparison of coordinate systems between PIC simulations and linear kinetic theory. We rotate the $x$-$z$ plane in linear kinetic theory about the $y$ axis to an orientation so that the wave vector of kinetic Alfv\'en wave is along the $x$ direction in PIC simulations.}
	\label{fig-coordinate}
\end{figure}

We carry out particle-in-cell (PIC) simulations using the OSIRIS framework \cite{fonseca2002osiris, hemker2015particle}, which consists of a massively parallel, fully relativistic, electromagnetic PIC code and a visualization and data analysis infrastructure. In this study, the simulations have one dimension ($x$) in configuration space and three dimensions ($v_x, v_y, v_z$) in velocity space. The cell length is $\Delta_x = \lambda_D$, where $\lambda_D = v_{Te} / \omega_{pe}$ is the initial electron Debye length, $v_{Te}$ is the initial electron thermal velocity, and $\omega_{pe}$ is the electron plasma frequency. The time step is set as $\Delta_t = 0.95 \Delta_x / c$ to satisfy the Courant-Friedrichs-Lewy condition in one dimension. The boundary conditions for both fields and particles are periodic. The coordinate system differs from what we used in linear kinetic theory: the $x$-$z$ plane is rotated with respect to the $y$ axis so that the wave propagates along the $x$ direction [Figure \ref{fig-coordinate}]. The background magnetic field $\mathbf{B}_0$ is oriented at a finite angle $\psi$ with respect to the $x$ axis in the $x$-$z$ plane. The electron cyclotron frequency $\omega_{ce}$ is equal to $0.2 \omega_{pe}$. Given the reduced ion-to-electron mass ratio $m_i / m_e = 64$, the normalized Alfv\'en velocity is $v_A / c = 0.2 \times \sqrt{\frac{1}{64}} = 0.025$. Both Ions and electrons are initialized as isotropic Maxwellian distributions. The electron and ion thermal velocities, $v_{Te}$ and $v_{Ti}$ respectively, are scaled to the Alfv\'en velocity $v_A$ to represent typical conditions in the equatorial magnetosphere \cite<e.g.,>{chaston2014observations}. To observe instabilities induced by nonlinear electron trapping in kinetic Alfv\'en waves, we need to reduce the background field fluctuations to a low level compared to the Alfv\'en wave field. For this purpose, each cell contains at least $10^5$ particles per species. Such computational cost is currently not affordable in $2$D and $3$D simulations, which is further reason why simulations are restricted to 1D in the present study.

We set up the Alfv\'en wave field by driving the plasma with an external pump field for a prescribed time interval. During this time interval, each particle experiences an external acceleration given by
\begin{linenomath*}
\begin{equation}
\left( \frac{d v_j}{d t} \right)_{\text{pump}} = \frac{q_s}{m_s} \operatorname{Re} \left\lbrace E_j e^{i k_0 x - i \omega_0 t} \right\rbrace,\, \text{with}\,\, j = x, y, z\,\, \text{and}\,\, s = e, i ,
\end{equation}
\end{linenomath*}
where $v_j$ is the particle velocity, $E_j$ is the pump electric field, and $t$ is time. $q_s$ and $m_s$ are the charge and mass of species $s$, respectively. The wavenumber and frequency of the pump field is $k_0$ and $\omega_0$, respectively. $k_0$ is connected to the mode number $M$ through $k_0 = 2 \pi M / (N_x \Delta_x)$, meaning the pump field has $M$ wavelengths in the system, where $N_x$ is the number of cells in the system. We choose the mode number $M = 2$ in our simulations. For a given $k_0$, the frequency $\omega_0$ is determined by the dispersion relation of kinetic Alfv\'en waves. We add the pump electric field to the self-generated electric field as the total electric field, and add the background magnetic field to the self-generated magnetic field as the total magnetic field. The total electric and magnetic fields are used in the particle push stage. The magnetic field associated with the kinetic Alfv\'en wave is generated naturally by the particle response. The time profile of the pump electric field is
\begin{linenomath*}
	\begin{equation}
	E_{j} = \begin{cases}
	E_{j 0} \cdot (t/t_{\text{rmp}}),\, & 0 \leqslant t < t_{\text{rmp}} \\
	E_{j 0},\, & t_{\text{rmp}} \leqslant t < t_{\text{off}} - t_{\text{rmp}} \\
	E_{j 0} \cdot [(t_{\text{off}} - t)/t_{\text{rmp}}],\, & t_{\text{off}} - t_{\text{rmp}} \leqslant t < t_{\text{off}} \\
	0,\, & t_{\text{off}} \leqslant t \leqslant t_{\text{end}}
	\end{cases}.
	\end{equation}
\end{linenomath*}
This pump field starts with a linear up-ramp until $t=t_{\text{rmp}}$, then maintains a constant amplitude until $t = t_{\text{off}} - t_{\text{rmp}}$, and finally ends with a linear down-ramp until $t = t_{\text{off}}$. The relative magnitude of the pump field, i.e., $E_{x0} / E_{y0}$ and $E_{z0} / E_{y0}$, is determined by the dispersion relation of kinetic Alfv\'en waves similar to that in Section \ref{sec-linear-kinetic}. The magnitude and duration of the pump field is chosen so that the field component $\delta B_y / B_0$ of the kinetic Alfv\'en wave reaches $\lesssim 0.1$ when the pump field is turned off. Such a large amplitude wave is needed to overcome the incoherent field fluctuations in the simulations. After the pump field is turned off, the electromagnetic field of kinetic Alfv\'en wave continues to propagate, and is self-consistently supported by the electron and ion distributions.

Below, we present two nominal simulations using the above setup (see detailed simulation parameters in Table \ref{tab-simulation-parameters}). In the first simulation ($\psi = 30^\circ$), the wave magnetic mirror force is a fraction of the parallel electric force $\vert F_{zb} / F_{ze} \vert = 0.2$, and these two forces are out of phase (i.e., $\arctan(F_{zb} / F_{ze}) = 180^\circ$), representing the regime of weak nonlinear Landau resonant interaction. In contrast, in the second simulation ($\psi = 75^\circ$), the wave magnetic mirror force is negligible (i.e., $\vert F_{zb} / F_{ze} \vert = 0.008$), representing the regime of strong nonlinear Landau resonant interaction. In both simulations, trapped electrons in kinetic Alfv\'en waves form electron beams and generate various forms of TDSs through beam instabilities.

\begin{table}[tphb]
	\caption{\label{tab-simulation-parameters}The parameters for two nominal simulations. The electric field is given in units of $\frac{m_e c^2}{e \cdot c / \omega_{pe}}$. The imaginary unit $i$ in the values of electric field represents a phase shift of $90$ degrees. It is surprising to find that the kinetic Alfv\'en wave at $\psi = 75^\circ$ is elliptically polarized but rotating in the electron sense, which was discovered by \citeA{gary1986low}.}
	\centering
	\begin{tabular}{l *{7}{c}}
		\hline
		               & $N_x$   & $\Delta_x / \lambda_D$  & $v_{Te} / v_A$   &  $v_{Ti} / v_A$ &  $M$   & $k_0 c / \omega_{pi}$  &  $\omega_0 / \omega_{ci}$ \\
		\hline
		Simulation 1   & $3561$  & $1$                       & $4.0$            &  $0.125$  &  $2$   & $0.28$                 &  $0.23$ \\
		Simulation 2   & $2059$  & $1$                      & $2.0$            &  $0.125$  &  $2$   & $0.98$                 &  $0.25$ \\
		\hline
		               & $\psi$      & $E_{x0}\, [\times 10^{-5}]$    & $E_{y0}\, [\times 10^{-5}]$     & $E_{z0}\, [\times 10^{-5}]$  & $t_{\text{rmp}}\omega_{ci}$ & $t_{\text{off}}\omega_{ci}$ & $t_{\text{end}}\omega_{ci}$  \\
		\hline
	    Simulation 1   & $30^\circ$  & $4.08 i$    & $3.0$        & $-7.59 i$ & $27.5$                      & $82.5$                     & $250$   \\
	    Simulation 2   & $75^\circ$  & $-11.19 i$  & $0.05$       & $3.17 i$  & $25.1$                      & $75.4$                     & $175$   \\
	    \hline
	\end{tabular}
\end{table}

\subsection{\label{sec-weak-landau}Weak nonlinear Landau resonant interaction}
\begin{figure}[tphb]
	\centering
	\includegraphics[width=4.5in]{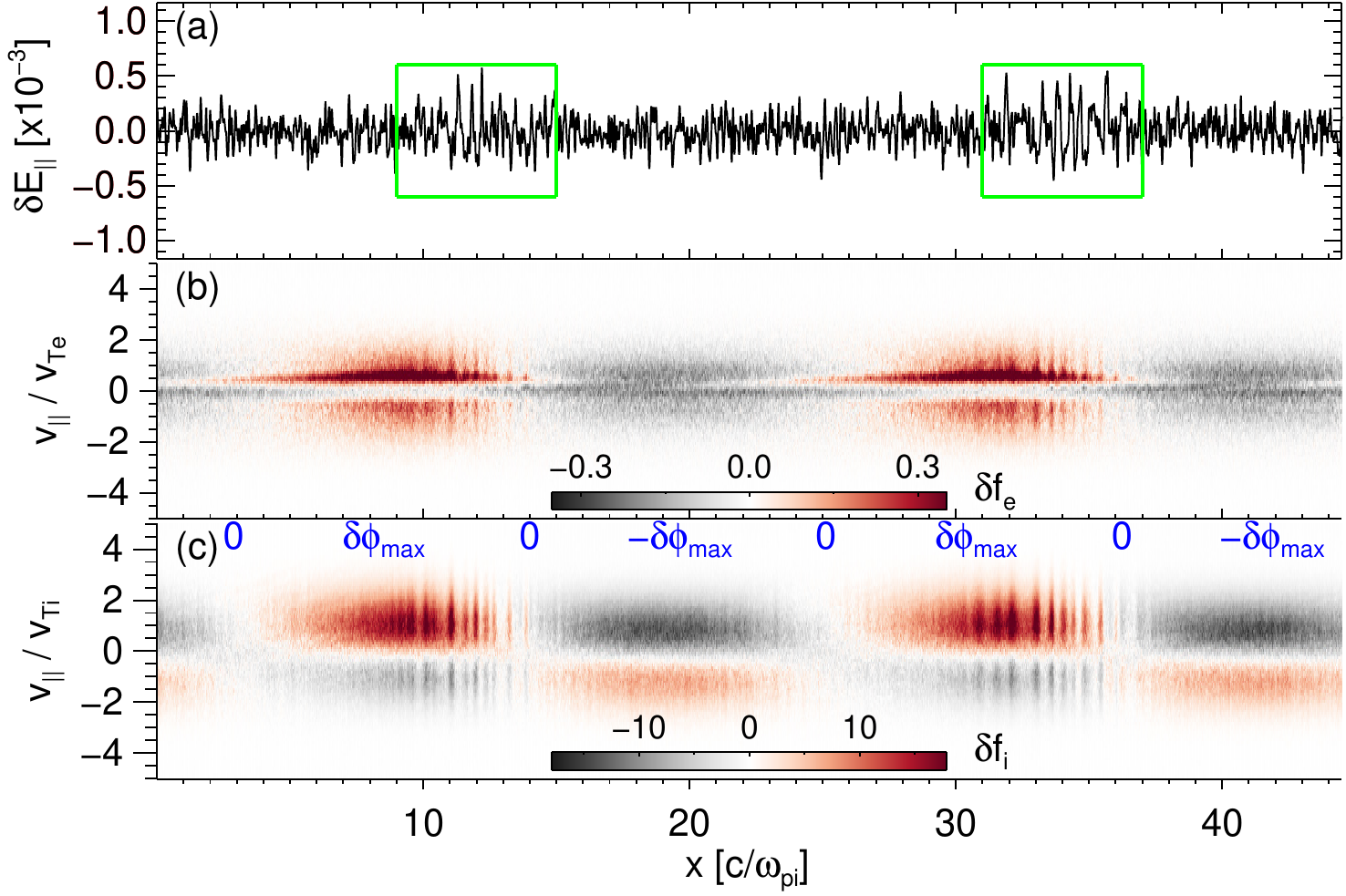}
	\caption{TDSs and phase space portraits in Simulation 1. This snapshot is taken at $t = 100\, \omega_{ci}^{-1}$. (a) The parallel electric field. The green boxes mark the TDS locations. \add{Hereafter in our PIC simulations, the electric field has the dimension $\frac{m_e c^2}{e \cdot c / \omega_{pe}}$.} (b) The perturbed electron phase space density $\delta f_e$. (c) The perturbed ion phase space density $\delta f_i$. At the top of this panel, the phase of the potential field is annotated by blue text.}
	\label{fig-ex-df-1}
\end{figure}

In the first simulation, the Landau resonant velocity, $\omega_0 / k_{0 \parallel} \simeq v_A$, is located at the core of the electron distribution ($v_A = 0.25 v_{Te}$) but at the tail of the ion distribution ($v_A = 8 v_{Ti}$). Therefore it is relatively easy to first understand the response of the ion distribution, since the core of ion distribution is characterized by a linear, non-resonant response [Figure \ref{fig-ex-df-1}(c)]. The linearized Vlasov equation for the reduced ion distribution (see \ref{append-reduce-lin-vlasov} for details of derivation) is
\begin{linenomath*}
	\begin{equation}\label{eq-linear-vlasov}
		\left( \frac{\partial}{\partial t} + v_{\parallel} \frac{\partial}{\partial x_\parallel} \right) \delta f_i = - \frac{q_i \delta E_\parallel}{m_i} \frac{\partial f_{0 i}}{\partial v_{\parallel}} ,
	\end{equation}
\end{linenomath*}
where $\delta f_i$ and $f_{0 i}$ are the perturbed and equilibrium parts of the reduced ion distribution function, respectively. Noticing $\delta E_\parallel \propto e^{i k_{0\parallel} x_\parallel - i \omega_0 t}$, we can Fourier analyze Equation \eqref{eq-linear-vlasov} and obtain
\begin{linenomath*}
	\begin{equation}\label{eq-ion-response}
		\delta f_i = -\frac{1}{\frac{\omega_0}{k_{0\parallel}} - v_\parallel} \frac{q_i \delta \phi}{m_i} \frac{\partial f_{0 i}}{\partial v_{\parallel}} ,
	\end{equation}
\end{linenomath*}
where $\delta \phi = \delta E_\parallel / (-i k_{0\parallel})$ is the potential of the kinetic Alfv\'en wave along the background magnetic field. Note that $\omega_0 / k_{0\parallel} > v_\parallel$ for the bulk of  the ion distribution. Thus the sign of the perturbation $\delta f_i$ depends on the sign of the potential field $\delta \phi$ and the velocity gradient $\partial f_{0i} / \partial v_\parallel$. Given a location $x$ in the ion phase space, the sign of $\delta f_i$ changes across $v_\parallel = 0$ due to the change of the sign in the velocity gradient $\partial f_{0i} / \partial v_\parallel$ [Figure \ref{fig-ex-df-1}(c)]. Conversely, we can infer the phase of the potential field $\delta \phi$ based on the ion perturbation $\delta f_i$, as annotated at the top of Figure \ref{fig-ex-df-1}(c). Knowledge of $\delta \phi$ will aid our analysis of electron phase trapping.

\begin{figure}[tphb]
	\centering
	\includegraphics[width=4.5in]{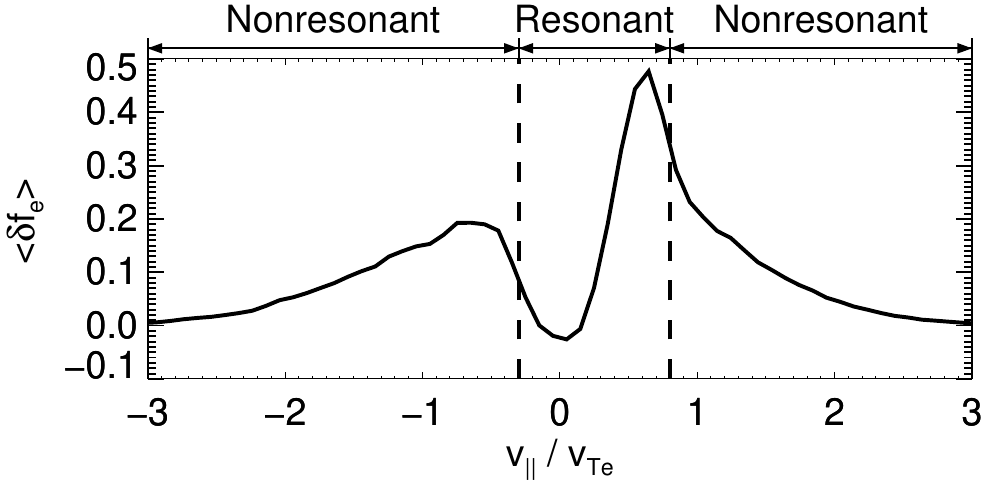}
	\caption{The localized electron velocity distribution function. This distribution is obtained by averaging the perturbed electron phase space density $\delta f_e$ in Figure \ref{fig-ex-df-1} between $x \omega_{pi} / c = 9$ and $x \omega_{pi} / c = 11$. The two vertical dashed lines indicate the extent of the separatrix around the trapping island, which separates resonant electrons from nonresonant electrons.}
	\label{fig-dfe-1dcut-1}
\end{figure}

The electron response to the kinetic Alfv\'en wave is mainly characterized by the formation of spatially modulated (or localized) beams around Landau resonance [Figure \ref{fig-ex-df-1}(b)]. The resonant electrons are accelerated in the phase of $\delta E_\parallel < 0$ (or $\partial \delta \phi / \partial x > 0$), whereas they are decelerated in the phase of $\delta E_\parallel > 0$ (or $\partial \delta \phi / \partial x < 0$). This transport of phase space density gives a spatially modulated beam distribution, which is centered around $v_A (= 0.25v_{Te})$ in velocity and peaked around $\delta \phi_{\max}$ in phase, known as the trapping island [Figure \ref{fig-ex-df-1}(b)]. To estimate the size of the trapping island, we average the perturbed phase space density $\delta f_e$ in the range $9 \leqslant x \omega_{pi} / c \leqslant 11$ and plot the averaged distribution $\langle\delta f_e\rangle$ in Figure \ref{fig-dfe-1dcut-1}. The trapping island (i.e., resonant region) is identified between the two dashed lines and has a half width $\Delta v_{tr} / v_{Te} = 0.55$. As a sanity check, we also calculate the half width of the trapping island using Equation \eqref{eq-vtr-practice} and obtain $\Delta v_{tr} / v_{Te} = 0.32$, where we have used the input $\delta B_z / B_0 = 0.013$ (obtained from the simulation) and $F_{zb} / F_{ze} = -0.2$ (derived from the linear kinetic theory). This theoretical estimation is roughly consistent with the result using simulation data.

Outside of the trapping island, the electrons are nonresonant and thus can be interpreted using
\begin{linenomath*}
	\begin{equation}\label{eq-nonres-electron-response}
	\delta f_e = \frac{1}{\frac{\omega_0}{k_{0\parallel}} - v_\parallel} \frac{e \delta \phi}{m_e} \frac{\partial f_{0 e}}{\partial v_{\parallel}} ,
	\end{equation}
\end{linenomath*}
which is on an equal footing with Equation \eqref{eq-ion-response} for the nonresonant ion response. With the use of Equation \eqref{eq-nonres-electron-response}, it is straightforward to demonstrate that a positive (negative) perturbation of electron phase space density $\delta f_e > 0$ ($\delta f_e < 0$) is produced in the phase of $\delta \phi > 0$ ($\delta \phi < 0$), as shown in Figure \ref{fig-ex-df-1}(b).

Electron beams driven by the kinetic Alfv\'en wave excite localized bursts of TDSs, appearing as bipolar electric field structures [Figure \ref{fig-ex-df-1}(a)]. \add{The ratio of the parallel electric field amplitude of TDSs to that of the kinetic Alfv\'en wave is about $17$.} These TDSs occur in the phase of $\delta E_{\parallel} > 0$ (or $\partial \delta \phi / \partial x < 0$). The exact occurrence phase of TDSs may depend on the cumulative growth rate $\int^t \gamma dt^\prime$ of beam instability over a certain time period, since the signal-to-noise ratio is modulated by $e^{{\int^t \gamma dt^\prime}}$. These TDSs are identified as nonlinear electron acoustic-mode \cite{holloway1991undamped, valentini2006excitation, anderegg2009electron}, which does not require the simultaneous presence of a cold electron component and a hot electron component as the usual electron acoustic-mode \cite{gary1993theory}. Instead, it survives undamped on the distribution of trapped electrons \cite{holloway1991undamped}. \add{In the space environment, the existence of kinetic Alfv\'en waves of finite amplitude indicates that a plateau of finite width on the electron distribution function has to be created, otherwise kinetic Alfv\'en waves would be damped out. As a result, TDSs survive in the space enviroment because their phase velocities are located within this plateau.}

\begin{figure*}[tphb]
	\centering
	\includegraphics{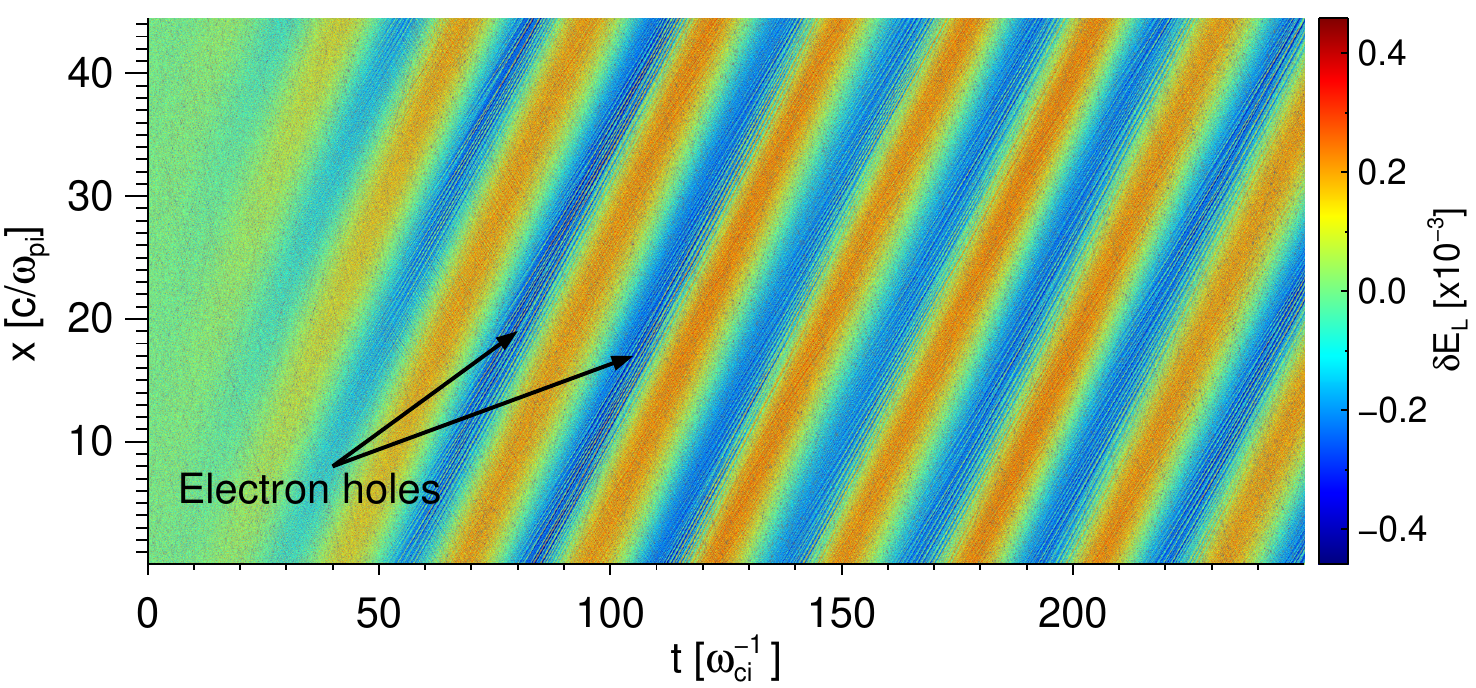}
	\caption{The spatiotemporal evolution of the longitudinal electric field $\delta E_L$ in Simulation 1. $\delta E_L$ is the projection of total electric field along the $\mathbf{k}$ direction. The phase and amplitude of $\delta E_L$ is color-coded.}
	\label{fig-wave-analysis-11}
\end{figure*}

\begin{figure}[tphb]
	\centering
	\includegraphics[width=4.5in]{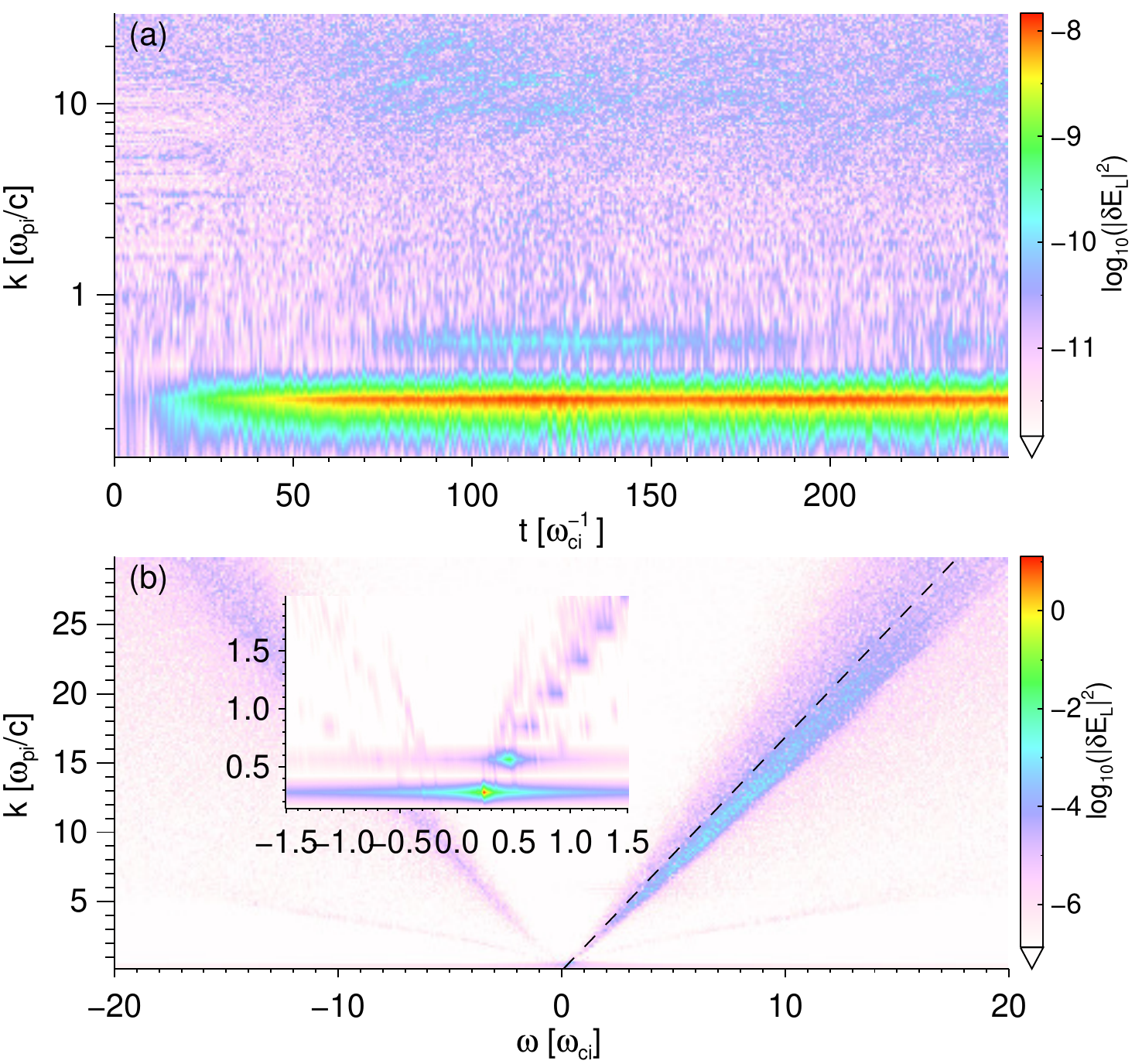}
	\caption{The Fourier spectra of the longitudinal electric field $\delta E_L$ in Simulation 1. (a) The temporal evolution of the wavenumber spectrum of $\delta E_L$. The wavenumber spectrum is obtained by Fourier-transforming the spatiotemporal pattern of $\delta E_L$ in space. The wavenumber of the fundamental kinetic Alfv\'en wave is $0.28 \omega_{pi} / c$. The wavenumber of TDS is in the range $10$ - $20\, \omega_{pi} / c$. (b) The frequency-wavenumber diagram of $\delta E_L$. The reciprocal of the slope of the dashed line represents the phase velocity of the kinetic Alfv\'en wave. \add{The inset plot zooms in on the fundamental kinetic Alfv\'en wave and its harmonics.} \add{We note that the weak signals propagating at $v = -v_A$ are likely caused by part of the resonant island extending to $v = -v_A$ as shown} in Figure \ref{fig-dfe-1dcut-1}.}
	\label{fig-wave-analysis-12}
\end{figure}

The electric field perturbation of TDSs is largely along the $\mathbf{k}$ direction (i.e., the longitudinal electric field $\delta E_L$). To analyze the TDS properties, we show the spatiotemporal evolution of $\delta E_L$ and its Fourier spectra in Figures \ref{fig-wave-analysis-11} and \ref{fig-wave-analysis-12}, respectively. TDSs propagate at a slightly larger phase velocity than the kinetic Alfv\'en wave [Figures \ref{fig-wave-analysis-11} and \ref{fig-wave-analysis-12}(b)]. The spiky electric field of TDSs has the signature of broadband spectrum [Figures \ref{fig-wave-analysis-12}]. Using the range of TDS wavenumbers in the spectrum, we can estimate the spatial scales of TDSs as $25 - 50 \lambda_D$. Due to the modest spatial bunching of trapped electrons, a weak second harmonic of the fundamental kinetic Alfv\'en wave is also generated [Figure \ref{fig-wave-analysis-12}(a)].


\begin{figure}[tphb]
	\centering
	\includegraphics[width=4.5in]{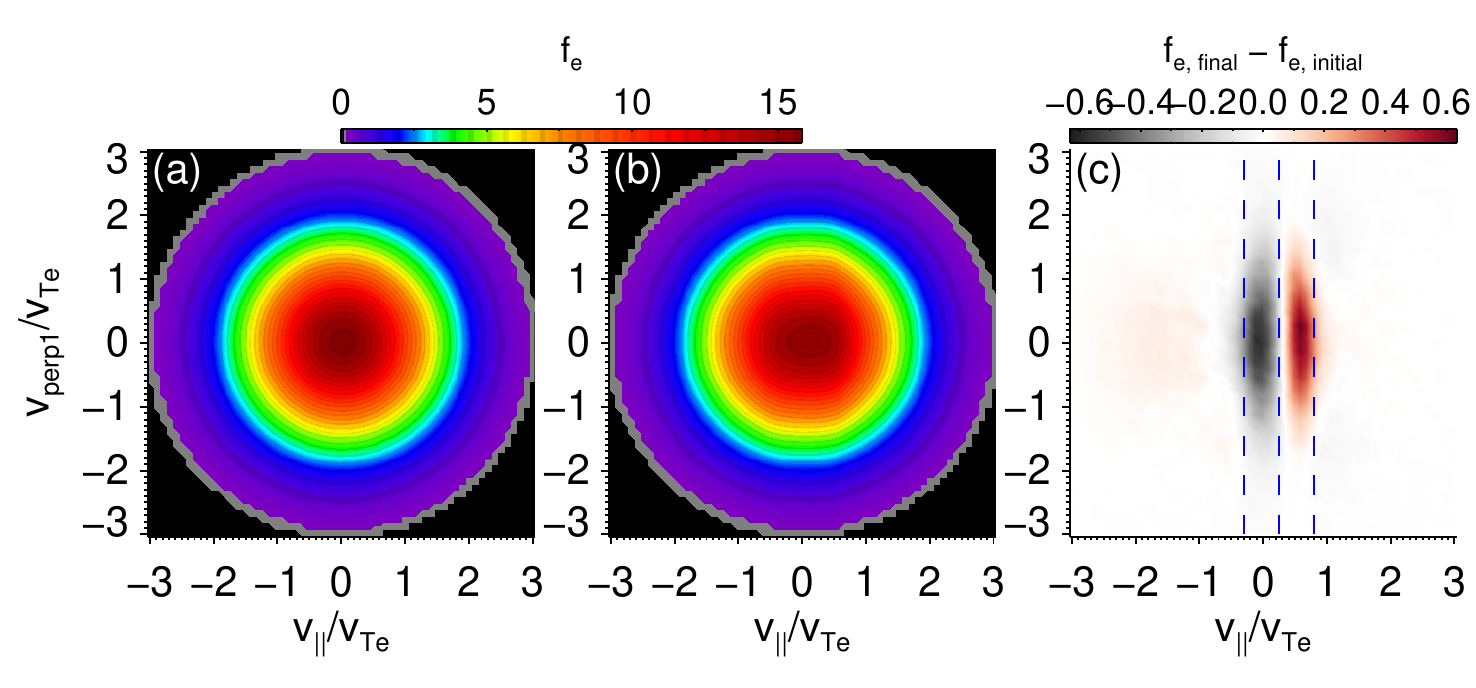}
	\caption{The thermal electron heating in Simulation 1. The density distribution in this figure has been averaged over the spatial domain. (a) The initial Maxwellian distribution. (b) The distribution at the end of the simulation. (c) The difference between the final and initial distribution functions. The three dashed lines from left to right stand for $v_\parallel = v_A - \Delta v_{tr}$, $v_\parallel = v_A$ and $v_\parallel = v_A + \Delta v_{tr}$, respectively.}
	\label{fig-eheating-1}
\end{figure}

The excitation of TDSs is a manifestation of beam instabilities driven by the kinetic Alfv\'en wave. The majority of wave energy, however, is deposited into thermal electrons through nonlinear Landau resonance. Figure \ref{fig-eheating-1} shows how the final electron distribution has deviated from the initial Maxwellian. Around the Landau resonant velocity $v_A$, a region of high phase space density moves from $v_\parallel < v_A$ to $v_\parallel > v_A$ and a region of low phase space density moves from $v_\parallel > v_A$ to $v_\parallel < v_A$ via nonlinear phase trapping. This results in a net increase in the kinetic energy of the resonant electrons and a consequent damping of the kinetic Alfv\'en wave. Again, the estimated half width of the trapping region using Equation \eqref{eq-vtr-practice}, $\Delta v_{tr} / v_{Te} = 0.55$, is consistent with that shown in Figure \ref{fig-eheating-1}(c).

\subsection{\label{sec-strong-landau}Strong nonlinear Landau resonant interaction}
In the second simulation, the wave magnetic mirror force is much smaller than the parallel electric force. The resulting half width of the trapping island is $\Delta v_{tr} / v_{Te} = 1$, where we have used Equation \eqref{eq-vtr-practice} with the input $\delta B_z / B_0 = 0.004$ (obtained from the simulation) and $F_{zb} / F_{ze} = 0.008$ (derived from the linear kinetic theory). For comparison, $\Delta v_{tr} / v_{Te}$ is about $0.55$ in the first simulation. Thus we would expect a stronger beam instability and potentially different characteristics of TDSs in the present case. Below we emphasize the differences between the first and second simulations.

\begin{figure}[tphb]
	\centering
	\includegraphics[width=4.5in]{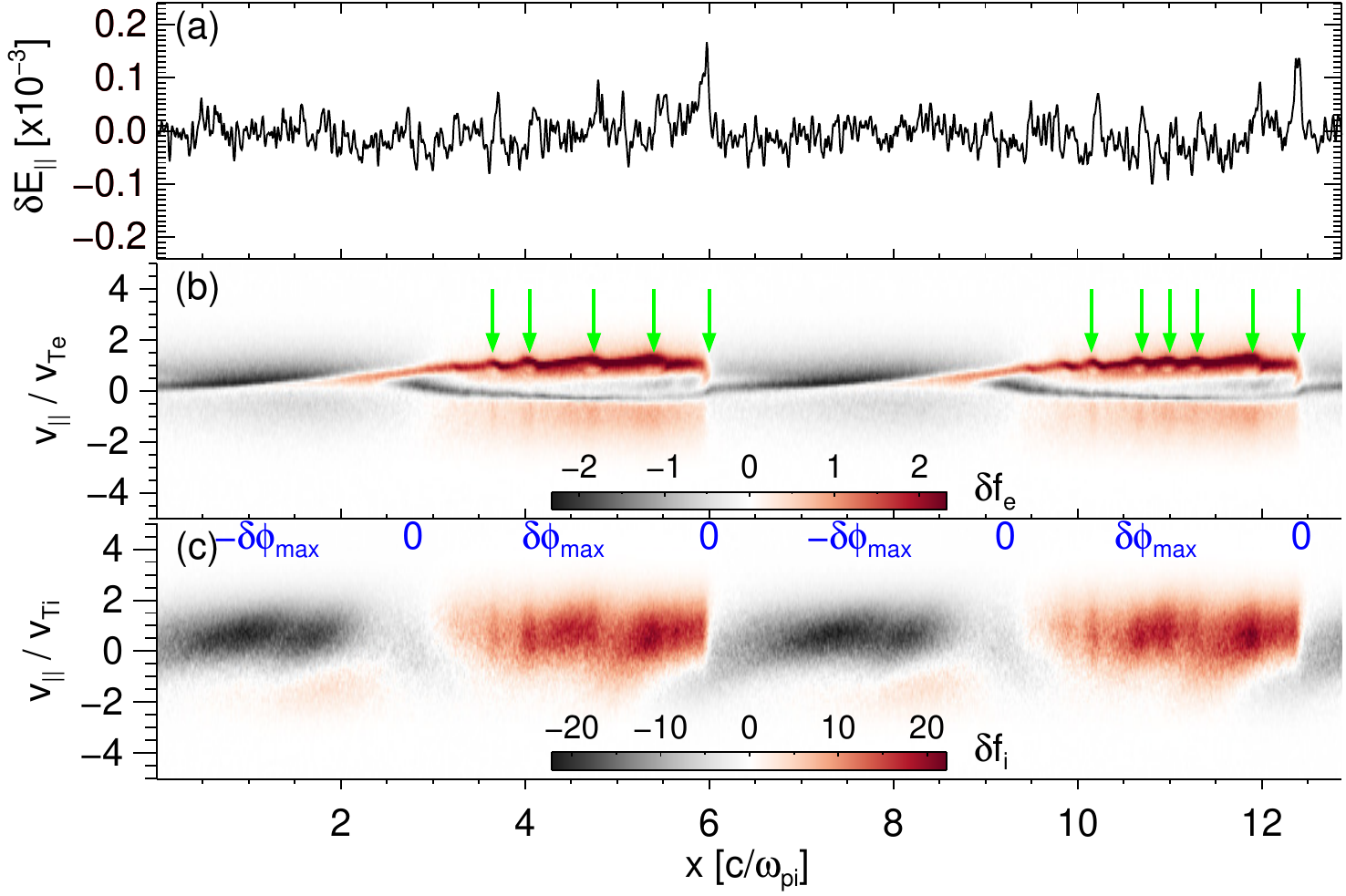}
	\caption{TDSs and phase space portraits in Simulation 2. This snapshot is taken at $t=61\, \omega_{ci}^{-1}$. (a) The parallel electric field. (b) The perturbed phase space density of electrons. Phase space structures in association with solitary electric fields are identified using arrows. Two of the arrows point to double layers at $x = 6.0\,\, \text{and}\,\, 12.5\, c / \omega_{pi}$. Other arrows point to phase space holes. (c) The perturbed phase space density of ions. The phase of potential field $\delta \phi$ is annotated by blue text.}
	\label{fig-ex-df-2}
\end{figure}

\begin{figure*}[tphb]
	\centering
	\includegraphics{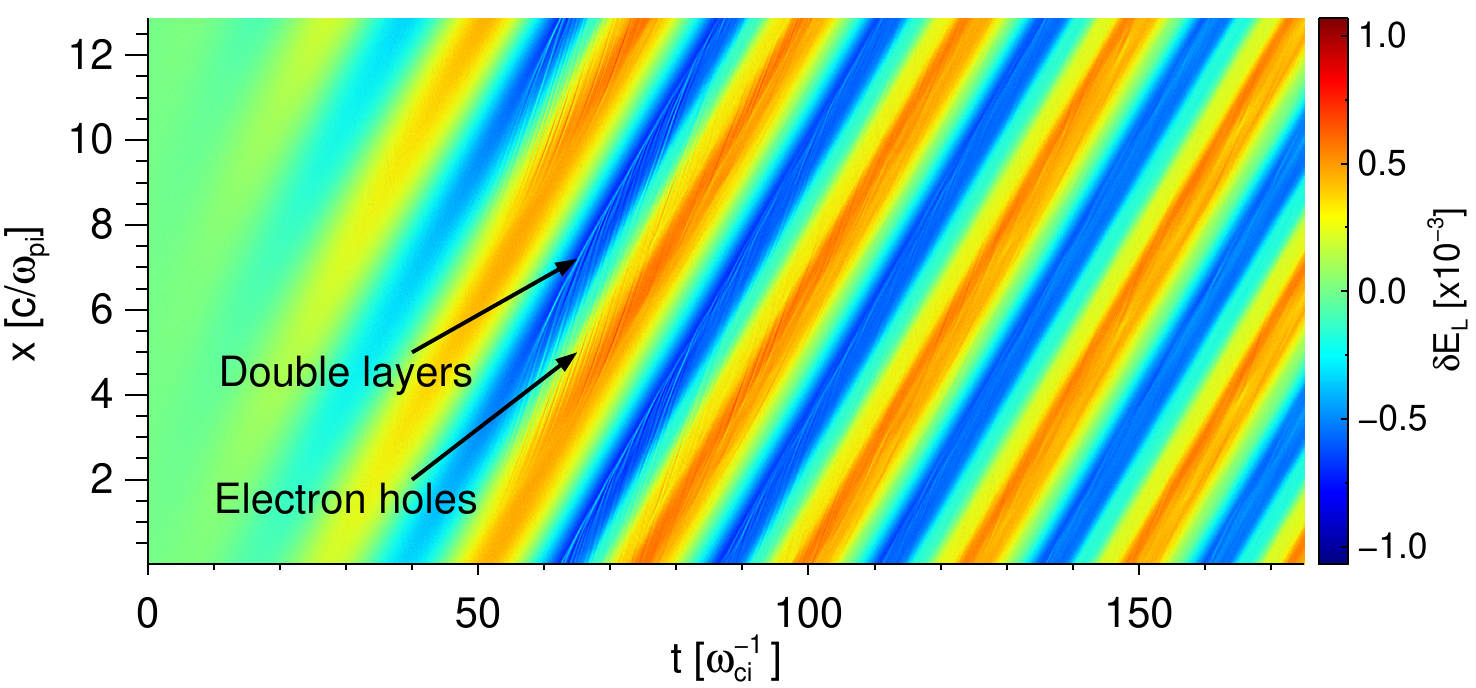}
	\caption{The spatiotemporal evolution of the longitudinal electric field $\delta E_L$ in Simulation 2. As indicated by the two arrows, phase space holes and double layers are embedded in the large scale kinetic Alfv\'en wave. The kinetic Alfv\'en wave field is distorted by its higher harmonics.}
	\label{fig-wave-analysis-21}
\end{figure*}

\begin{figure}[tphb]
	\centering
	\includegraphics[width=4.5in]{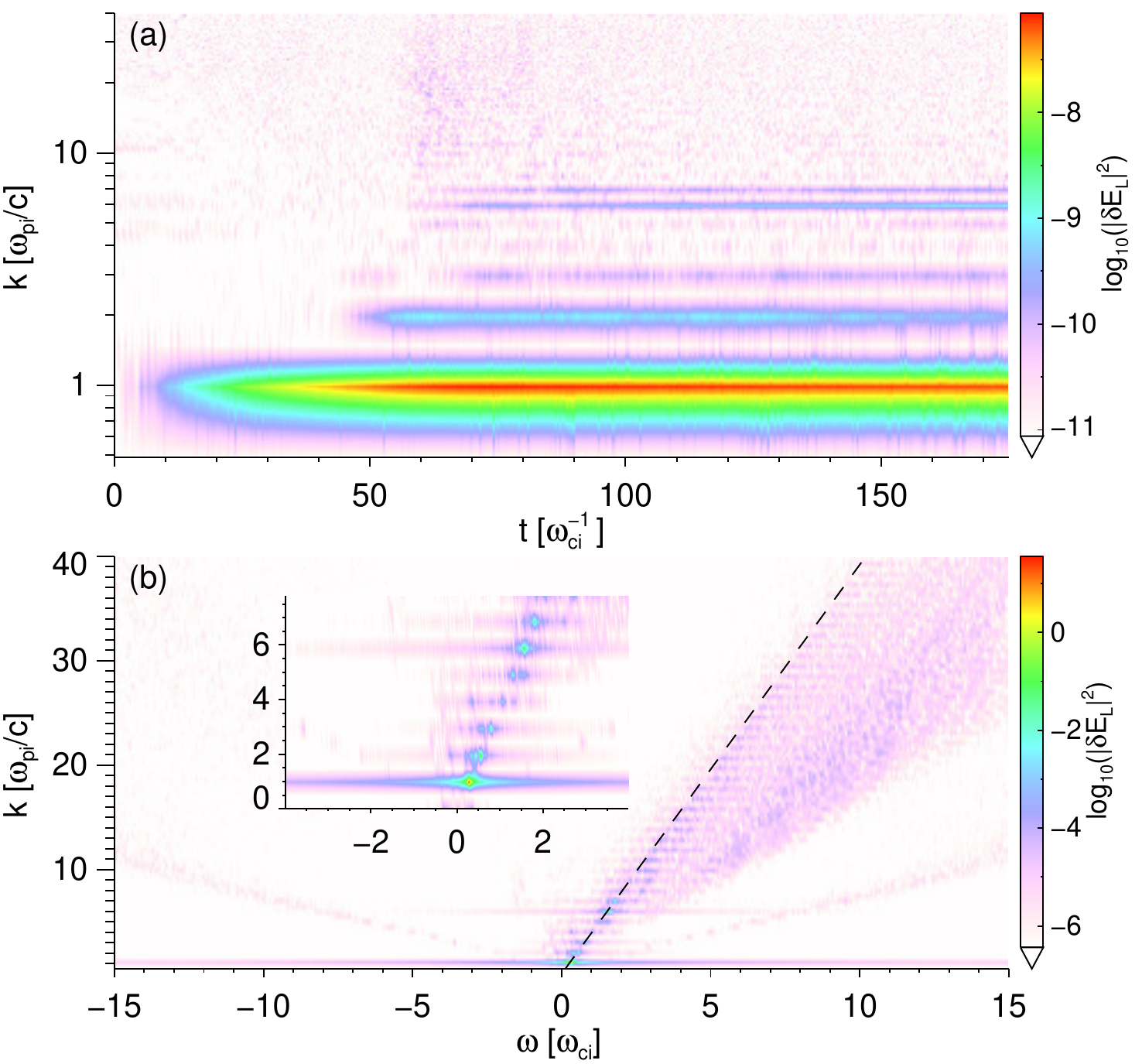}
	\caption{The Fourier spectra of the longitudinal electric field $\delta E_L$ in Simulation 2. (a) The temporal evolution of the wavenumber spectrum of $\delta E_L$. The wavenumber of the fundamental kinetic Alfv\'en wave is $0.98 \omega_{pi} / c$. Higher harmonics of the kinetic Alfv\'en wave occur consecutively in time. TDSs start to be excited in the interval $50 < t \omega_{ci} < 100$ and are mainly located in the wavenumber range $10$ - $30\, \omega_{pi} / c$. (b) The frequency-wavenumber diagram of $\delta E_L$. Notably, the propagation velocity of TDSs is greater than that of the kinetic Alfv\'en wave, whereas the harmonics of the kinetic Alfv\'en wave propagate at the same velocity as the fundamental mode. \add{The inset plot zooms in on the fundamental kinetic Alfv\'en wave and its harmonics.}}
	\label{fig-wave-analysis-22}
\end{figure}

The expectation of a strong beam instability is confirmed by the electron phase space plot [Figure \ref{fig-ex-df-2}(b)]. Spatially modulated, prominent electron beams are driven by the parallel electric field of the kinetic Alfv\'en wave inside the trapping island ($v_A - \Delta v_{tr} < v_\parallel < v_A + \Delta v_{tr}$). Solitary electric field structures are generated by these unstable beams. Several phase space holes are clearly identified in Figures \ref{fig-ex-df-2}(a) and \ref{fig-ex-df-2}(b). \add{The ratio of the parallel electric field amplitude of TDSs to that of the kinetic Alfv\'en wave is about $10$.} The phase space holes propagate at the local beam velocity, which is larger than $v_A$ [Figures \ref{fig-ex-df-2}(b) and \ref{fig-wave-analysis-22}(b)]. As a consequence, the phase space holes overtake the phase fronts of the kinetic Alfv\'en wave [Figure \ref{fig-wave-analysis-21}]. In addition, double layers are seen to form in the phase of maximum $\delta E_{\parallel}$, where the phase of $\delta E_{\parallel}$ is inferred from $\delta f_i$ using the technique in Section \ref{sec-weak-landau}. The beam electrons are slowed down by the double layers and accumulate at the high potential sides of the double layers [Figure \ref{fig-ex-df-2}(b)], which eventually leads to the dissipation of double layers [Figure \ref{fig-wave-analysis-21}]. Finally, in comparison with the first simulation, more harmonics of the kinetic Alfv\'en wave are generated due to the nonlinear phase trapping of electrons [Figures \ref{fig-wave-analysis-21} and \ref{fig-wave-analysis-22}].

\begin{figure}[tphb]
	\centering
	\includegraphics[width=4.5in]{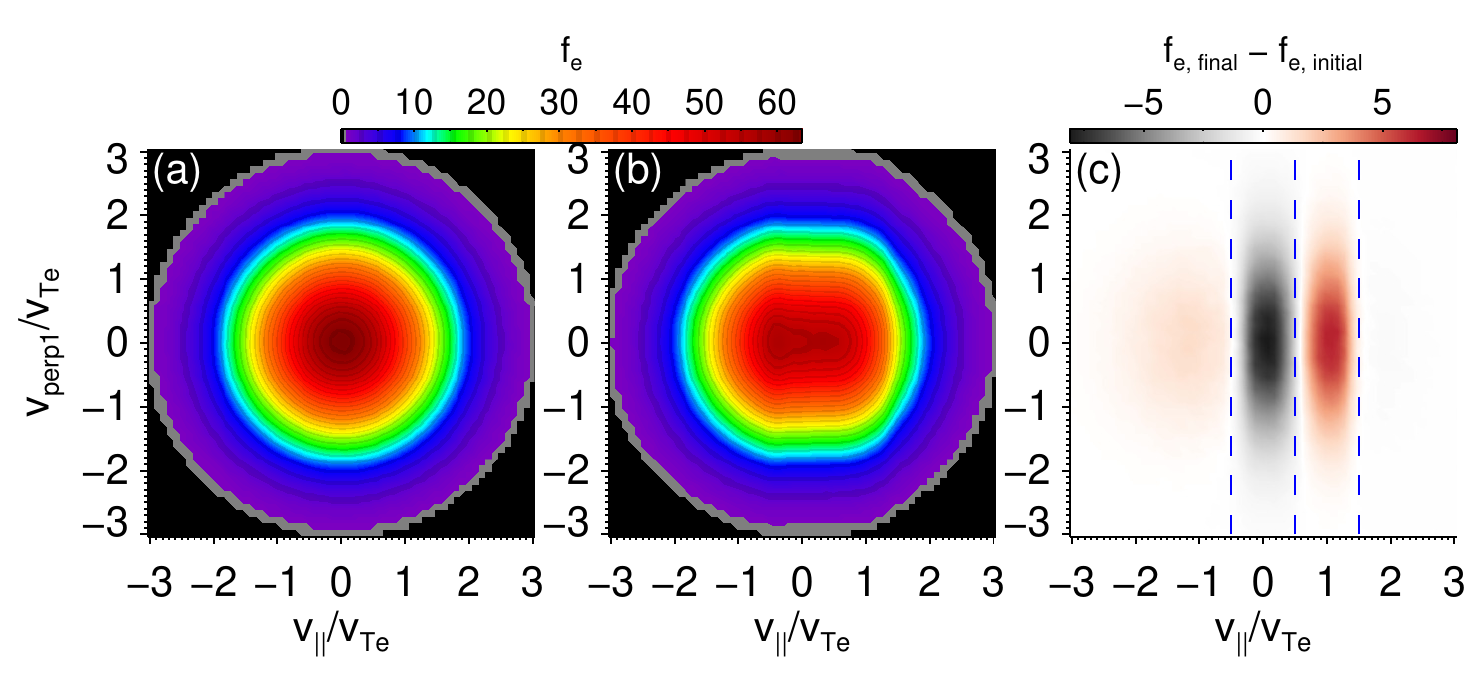}
	\caption{The thermal electron heating in Simulation 2. The density distribution has been averaged over the spatial domain. (a) The initial Maxwellian distribution. (b) The final flat-top distribution. (c) The difference between the final and initial distribution functions. The three dashed lines from left to right stand for $v_\parallel = v_A - \Delta v_{tr}$, $v_\parallel = v_A$ and $v_\parallel = v_A + \Delta v_{tr}$, respectively.}
	\label{fig-eheating-2}
\end{figure}

Thermal electrons in the present simulation are more strongly heated through nonlinear Landau resonance than the first simulation. This produces an elongated electron distribution in the parallel velocity, known as the flat-top distribution [Figure \ref{fig-eheating-2}]. Such a distribution has been previously obtained in coordinated observations and simulations \cite<e.g.,>{damiano2018electron}. The velocity range of the flat-top is consistent with the Landau resonance range, $v_A - \Delta v_{tr} < v_\parallel < v_A + \Delta v_{tr}$. Curiously, weak non-resonant electron heating is seen next to the Landau resonance at $v_\parallel < 0$.

\subsection{Critical condition for TDS excitation}\label{sec-tds-excitation-condition}
As seen from the PIC simulations, the free energy source of TDSs is the trapped electron beam driven by the kinetic Alfv\'en wave. However, trapped electrons are subject to phase mixing \cite{o1965collisionless} and thus the beam distribution is destroyed in a few trapping periods. \add{Note that the phase mixing here refers to phase mixing of nonlinear Landau resonant electrons of different energies}. How does the phase mixing rate of trapped electrons compare with the growth rate of the beam instability? To address this problem, we analyze the controlling factors of this process.

\begin{figure}[tphb]
	\centering
	\includegraphics[width=4.5in]{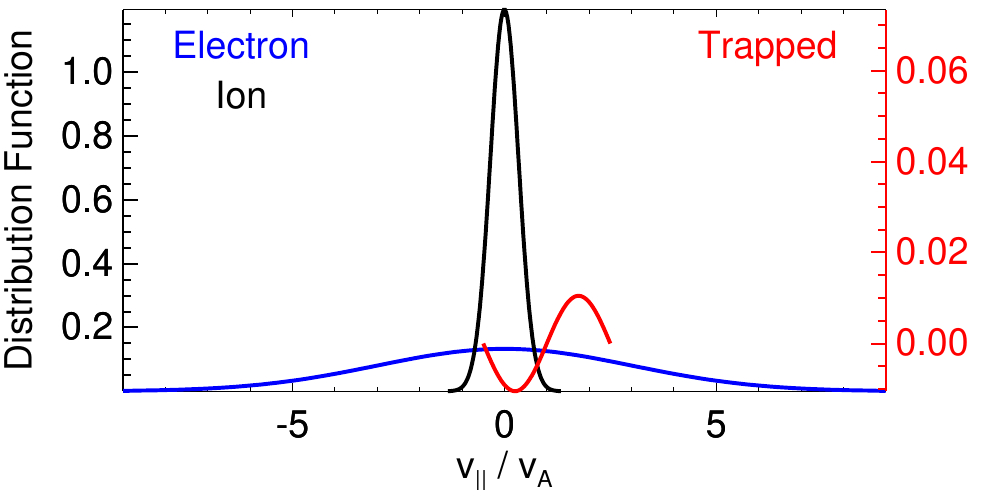}
	\caption{Sketch of distribution functions. The black line represents the ion Maxwellian distribution $f_{0i}$. The electron distribution is separated into an equilibrium part (blue; the Maxwellian distribution $f_{0e}$) and a perturbed part [red; $\delta f_{tr}$ in Equation \eqref{eq-trapped-distribution}]. The perturbed part is induced by trapped electrons, the scale of which is shown by the vertical axis on the right.}
	\label{fig-distribution-control-factors}
\end{figure}

We separate the electron distribution into a Maxwellian $f_{0e} = \frac{1}{\sqrt{2 \pi} v_{Te}} \exp\left(-\frac{v_\parallel^2}{2 v_{Te}^2}\right)$ (by initialization) and a perturbed distribution $\delta f_{tr}$ (by trapped electrons) [see Figure \ref{fig-distribution-control-factors}]. The perturbed distribution comprises of a dip in the range $v_A - \Delta v_{tr} < v_\parallel < v_A$ and a bump in the range $v_A < v_\parallel < v_A + \Delta v_{tr}$, which is formally modeled as
\begin{linenomath*}
	\begin{equation}\label{eq-trapped-distribution}
		\delta f_{tr} = \Delta f \cdot \sin\left(\pi \frac{v_\parallel - v_A}{\Delta v_{tr}}\right) .
	\end{equation}
\end{linenomath*}
The magnitude of the perturbed distribution, $\Delta f$, is determined by
\begin{linenomath*}
	\begin{equation}
	\Delta f = \frac{ f_{0e}(v_A - \Delta v_{tr} / 2) - f_{0e}(v_A + \Delta v_{tr} / 2) }{2} .
	\end{equation}
\end{linenomath*}
Expanding $f_{0e}$ about $v_\parallel = v_A$ in Taylor series, we can rewrite $\Delta f$ as
\begin{linenomath*}
	\begin{equation}\label{eq-Delf-Taylor-expansion}
	\begin{split}
	\Delta f &= - \sum\limits_{n=0}^{\infty} \frac{1}{(2n+1)!} \left(\frac{\Delta v_{tr}}{2}\right)^{2n+1} f_{0e}^{(2n+1)}(v_A) \\
    &= \sum\limits_{n=0}^{\infty} \frac{H_{2n+1}\left(\frac{v_A}{\sqrt{2}v_{Te}}\right)}{(2n+1)!} \left(\frac{\Delta v_{tr}}{2\sqrt{2}v_{Te}}\right)^{2n+1}  f_{0e}(v_A) ,
	\end{split}
	\end{equation}
\end{linenomath*}
where even terms in the Taylor series are canceled out and the Hermite Polynomial $H_{2n+1}$ is used for calculating derivatives of the Maxwellian distribution \cite{weber2003essential}.

The growth rate of the beam instability contributed by the trapped electrons can be written as \cite{o1968transition}
\begin{linenomath*}
	\begin{equation}\label{eq-beam-growth-rate}
	\begin{split}
	\frac{\gamma_{tr}}{k_\parallel v_{Te}} &= \frac{\pi}{2} \frac{\omega}{k_\parallel v_{Te}} \frac{\omega_{pe}^2}{k_\parallel^2} \left(\frac{\partial}{\partial v_{\parallel}} \delta f_{tr}\right)_{v_\parallel = \omega / k_{\parallel}} \\
	& \lesssim \frac{\pi}{2} \frac{\omega}{k_\parallel v_{Te}} \frac{\omega_{pe}^2}{k_\parallel^2} \left(\frac{\partial}{\partial v_{\parallel}} \delta f_{tr}\right)_{v_\parallel = v_A} \\
	& = \frac{\pi}{2} \frac{\omega}{k_\parallel v_{Te}} \frac{1}{k_\parallel^2 \lambda_D^2} \frac{\pi}{2 \sqrt{2}} v_{Te} f_{0e}(v_A) \\
	& \times \sum\limits_{n=0}^{\infty} \frac{H_{2n+1}\left(\frac{v_A}{\sqrt{2}v_{Te}}\right)}{(2n+1)!} \left(\frac{e \delta \phi}{2 T_e}\right)^{n}  ,
	\end{split}
	\end{equation}
\end{linenomath*}
where $\omega / k_\parallel$ is the phase velocity of TDSs. The zeroth order term ($n=0$) of the beam growth rate does not depend on the wave amplitude $\delta \phi$ and always provides a positive growth rate [$H_1(v_A / \sqrt{2}v_{Te}) > 0$]. The next higher order term ($n=1$) linearly scales with $\delta \phi$, the sign of which depends on $H_3(v_A / \sqrt{2}v_{Te})$. For the range of interest (i.e., $v_A < v_{Te}$), we have $H_3(v_A / \sqrt{2}v_{Te}) < 0$, which means the beam growth rate decreases with the wave amplitude $\delta \phi$ in the regime of $v_A < v_{Te}$.

In the meantime, TDSs undergo Landau damping caused by $f_{0e}$ and $f_{0i}$ [Figure \ref{fig-distribution-control-factors}]. The ratio of electron Landau damping rate to ion Landau damping rate is
\begin{linenomath*}
	\begin{equation}
	\frac{\gamma_{L e}}{\gamma_{L i}} = \frac{m_i}{m_e} \cdot \frac{v_{Ti}^3}{v_{Te}^3} e^{-\frac{v_A^2}{v_{Te}^2} + \frac{v_A^2}{v_{Ti}^2}} .
	\end{equation}
\end{linenomath*}
For typical values $T_e = 300$\,eV, $T_i = 2000$\,eV, $B_0 = 100$\,nT, $n_0 = 1\,\text{cm}^{-3}$ around dipolarization fronts, we have $\vert \gamma_{L e} \vert \gg \vert \gamma_{L i} \vert$. Thus it suffices to consider only the electron Landau damping rate, i.e.,
\begin{linenomath*}
	\begin{equation}\label{eq-landau-damping-rate}
	\begin{split}
	\frac{\gamma_{L e}}{k_\parallel v_{Te}} &= \frac{\pi}{2} \frac{\omega}{k_\parallel v_{Te}} \frac{\omega_{pe}^2}{k_\parallel^2} f_{0e}^\prime(v_A) \\
	&= - \frac{\pi}{2} \frac{\omega}{k_\parallel v_{Te}} \frac{1}{k_\parallel^2 \lambda_D^2} \frac{1}{\sqrt{2}} v_{Te} f_{0e}(v_A) H_1\left(\frac{v_A}{\sqrt{2}v_{Te}}\right) .
	\end{split}
	\end{equation}
\end{linenomath*}

\begin{figure}[tphb]
	\centering
	\includegraphics[width=4.5in]{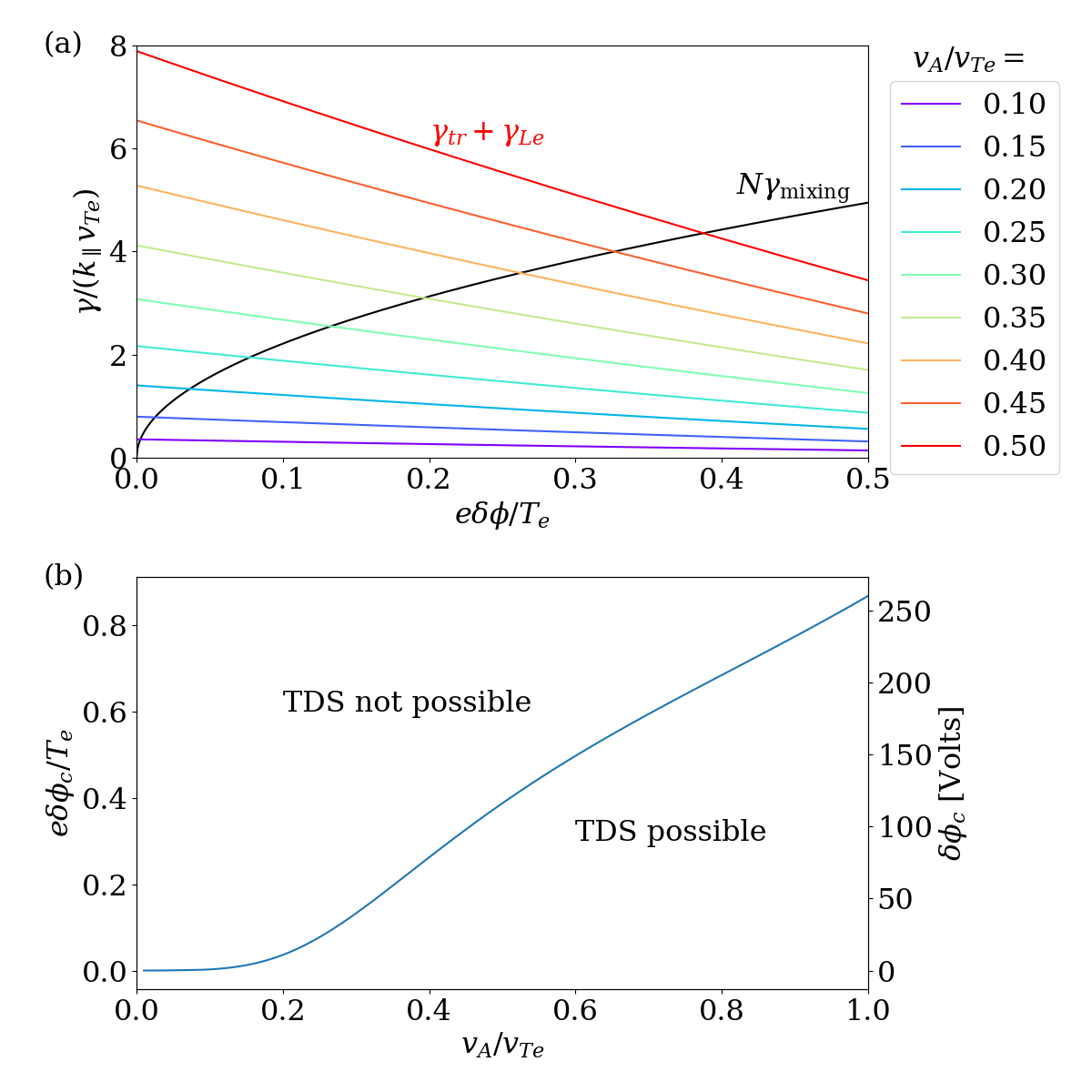}
	\caption{Critical condition for TDS excitation. (a) The comparison between the TDS growth rate and the phase mixing rate. In evaluating the TDS growth rate on the left side of Inequality \eqref{eq-excitation-condition}, we use the typical TDS parameter $k_\parallel \lambda_D = 0.1$, truncate the power series at $n=4$ (the resulting relative truncation error is less than $5 \times 10^{-5}$), and parameterize the TDS growth rate by $v_A / v_{Te}$. The results are shown in rainbow-colored lines. In evaluating the phase mixing rate on the right side of Inequality \eqref{eq-excitation-condition}, we take the number of e-folding as $N = 7$ (i.e., the signal-to-noise ratio is $e^{7} \simeq 10^3$). The result is shown in the black line. \add{For given values of $v_A / v_{Te}$ and $e \delta \phi / T_e$, if the phase mixing rate is greater than the TDS growth rate, excitation of TDSs is prohibited due to electron phase mixing.} (b) The critical potential amplitude of kinetic Alfv\'en waves to drive TDSs. Beyond the critical potential amplitude $\delta \phi_c$, TDSs cannot be excited through electron trapping by kinetic Alfv\'en waves. \add{The texts ``TDS possible'' and ``TDS not possible'' indicate the allowed and prohibited regions respectively in the parameter space for excitation of TDSs.} To have a better comparison with spacecraft observations, the critical potential $\delta \phi_c$ is shown in the physical unit ``volts'' on the $y$ axis on the right for a typical electron temperature $T_e = 300$\,eV.}
	\label{fig-critical-amplitude}
\end{figure}

The phase mixing rate of trapped electrons is characterized by the trapping frequency, i.e.,
\begin{linenomath*}
	\begin{equation}\label{eq-phase-mixing-rate}
		\frac{\gamma_{\text{mixing}}}{k_\parallel v_{Te}} \sim \frac{\omega_{tr}}{k_\parallel v_{Te}} = \sqrt{\frac{e \delta \phi}{T_e}} \propto \sqrt{\delta \phi} .
	\end{equation}
\end{linenomath*}
The signal-to-noise ratio of TDSs can be estimated as $e^{(\gamma_{tr} + \gamma_{L e}) \Delta t_{\text{mixing}}}$, where $\Delta t_{\text{mixing}} = 1 / \gamma_{\text{mixing}}$. Suppose that TDSs are observable after $N$ $e$-foldings. The critical condition for TDS excitation may be written as
\begin{linenomath}
	\begin{equation}\label{eq-control}
		\gamma_{tr} + \gamma_{L e} \geqslant N \gamma_{\text{mixing}} .
	\end{equation}
\end{linenomath}
Plugging Equations \eqref{eq-beam-growth-rate}, \eqref{eq-landau-damping-rate} and \eqref{eq-phase-mixing-rate} into Equation \eqref{eq-control}, we explicitly obtain
\begin{linenomath*}
	\begin{equation}\label{eq-excitation-condition}
	\begin{split}
	\frac{\pi}{2 \sqrt{2}} \frac{\omega}{k_\parallel v_{Te}} \frac{1}{k_\parallel^2 \lambda_D^2} v_{Te} f_{0e}(v_A) \left[ \frac{\pi}{2} \sum\limits_{n=0}^{\infty} \frac{H_{2n+1}\left(\frac{v_A}{\sqrt{2}v_{Te}}\right)}{(2n+1)!} \left(\frac{ e \delta \phi}{2 T_e}\right)^{n} - H_1\left(\frac{v_A}{\sqrt{2}v_{Te}}\right) \right] \geqslant N \sqrt{\frac{e \delta \phi}{T_e}} .
	\end{split}
	\end{equation}
\end{linenomath*}
The TDS growth rate [left side of Inequality \eqref{eq-excitation-condition}] decreases with $\delta \phi$, whereas the phase mixing rate [right side of Inequality \eqref{eq-excitation-condition}] increases as $\sqrt{\delta \phi}$. This gives an upper bound of wave amplitude $\delta \phi_c$, beyond which the phase mixing rate exceeds the TDS growth rate and thus TDSs cannot be excited anymore. To find $\delta \phi_c$, we evaluate both sides of Inequality \eqref{eq-excitation-condition} as shown in Figure \ref{fig-critical-amplitude}(a). In this procedure, we parameterize the TDS growth rate by $v_A / v_{Te}$. For a given normalized wave amplitude $e \delta \phi / T_e$, the TDS growth rate ($\gamma_{tr} + \gamma_{L e}$) increases with $v_A / v_{Te}$. For a given $v_A / v_{Te}$, the phase mixing rate exceeds the TDS growth rate at the critical amplitude $e \delta \phi_c / T_e$. We plot this critical amplitude $e \delta \phi_c / T_e$ as a function of $v_A / v_{Te}$ in Figure \ref{fig-critical-amplitude}(b). For example, the critical amplitude is $e \delta \phi_c / T_e = 0.078$ for $v_A / v_{Te} = 0.25$ in Simulation 1, whereas the critical amplitude is $e \delta \phi_c / T_e = 0.387$ for $v_A / v_{Te} = 0.5$ in Simulation 2. Using Equation \eqref{eq-trapping-island-width}, we calculate the critical half width of the trapping island as $\Delta v_{tr, c} / v_{Te} = 0.56$ for Simulation 1 and $\Delta v_{tr, c} / v_{Te} = 1.24$ for Simulation 2. The measured half widths of the trapping island are below these critical values in each simulation. Additionally, we have confirmed that the excitation of TDSs is prohibited if the wave amplitude is above the critical value $\delta \phi_c$ in PIC simulations.

\section{Conclusions}\label{sec-conclusions}
In this paper we have presented the excitation of TDSs through nonlinear Landau resonant interaction between kinetic Alfv\'en waves and thermal electrons. First, we show that the parallel electric field of the kinetic Alfv\'en wave is the primary driver of electron phase trapping. Second, we demonstrate that a spatially modulated beam distribution is formed by phase-trapped electrons and excites TDSs through the beam instability. Thermal electrons are heated by the kinetic Alfv\'en wave in the nonlinear trapping process. Third, we demonstrate that the TDS growth rate decreases with the wave potential $\delta \phi$ whereas the phase mixing rate scales with $\sqrt{\delta \phi}$. A critical condition for TDS excitation is thus derived [Equation \eqref{eq-excitation-condition}] and an upper bound of $\delta \phi$ is obtained [Figure \ref{fig-critical-amplitude}].

\begin{figure}[tphb]
	\centering
	\includegraphics[width=4.5in]{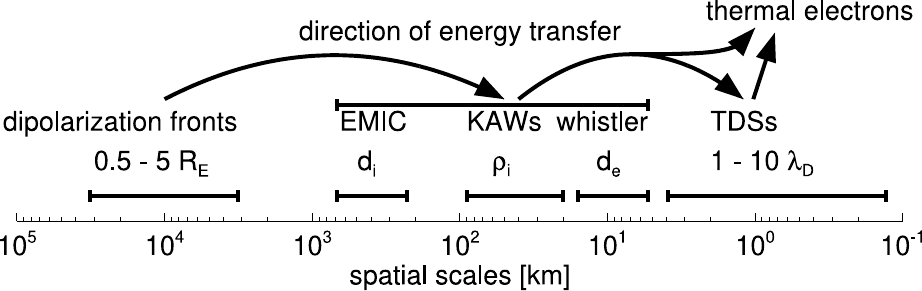}
	\caption{Energy cascading from macroscopic dipolarization fronts down to microscopic TDSs. In characterizing the spatial scales of different objects, we use the following typical parameters around dipolarization fronts: the background magnetic field $50$ - $150$\,nT; plasma density $0.1$ - $1\, \text{cm}^{-3}$; ion temperature $1000$ - $2000$\,eV; electron temperature $300$\,eV. ``KAWs'' is the acronym for kinetic Alfv\'en waves.}
	\label{fig-scale}
\end{figure}

Putting this work in the bigger context regarding the dissipation of the injection energy in the inner magnetosphere, a picture of energy cascading emerges as follows [Figure \ref{fig-scale}]: (1) The energy carried by the flow and the magnetic field of dipolarization fronts at macroscale is first converted to that of kinetic Alfv\'en waves and whistler waves; (2) These kinetic Alfv\'en waves and whistler waves then drive TDSs by accelerating electron beams locally through nonlinear Landau resonance; (3) In the meantime, kinetic Alfv\'en waves and whistler waves, together with TDSs, heat thermal electrons. In this way, energy cascades from dipolarization fronts at the macroscale to TDSs at the microscale, and is eventually deposited to electron thermal energy. Such a picture of energy cascading is being actively sought by combining global MHD simulations, test particle simulations and kinetic instability analysis to conquer the vast scale separations between dipolarization fronts and TDSs \cite<e.g.,>[]{ukhorskiy2018microscopic, ukhorskiy2019kinetic}.

\appendix
\section{\label{append-reduce-lin-vlasov}Linearized Vlasov equation for the reduced particle distribution}
We use the the linearized Vlasov equation to describe the nonresonant response of the particle distribution. This equation reads
\begin{linenomath*}
	\begin{equation}\label{eq-full-lin-vlasov}
		\left(\frac{\partial}{\partial t} + \mathbf{v} \cdot \frac{\partial}{\partial \mathbf{x}}\right) \delta F_{s} = - \frac{q_s}{m_s} \left[\delta \mathbf{E} + \frac{\mathbf{v} \times (\delta \mathbf{B} + \mathbf{B}_{0})}{c}\right] \cdot \frac{\partial F_{0s}}{\partial \mathbf{v}} ,
	\end{equation}
\end{linenomath*}
where $\delta F_s (t, \mathbf{x}, \mathbf{v})$ and $F_{0 s} (\mathbf{v})$ are the perturbed and equilibrium parts of the distribution function of species $s$, respectively. We aim to rewrite the linearized Vlasov equation in terms of the reduced distribution functions
\begin{linenomath*}
	\begin{eqnarray}
		\delta f_s &=& \iint dv_1\, dv_2\, \delta F_s = \langle \delta F_s \rangle , \\
		f_{0 s}    &=& \iint dv_1\, dv_2\, F_{0 s} = \langle F_{0 s} \rangle ,
	\end{eqnarray}
\end{linenomath*}
where $v_1$ and $v_2$ are the two orthogonal velocity components perpendicular to $\mathbf{B}_0 = B_0 \hat{\mathbf{x}}_3$, and the notation $\langle \cdot \rangle$ is short for the velocity space integral $\iint dv_1\, dv_2$. By integrating Equation \eqref{eq-full-lin-vlasov} over $v_1$ and $v_2$, we obtain
\begin{linenomath*}
	\begin{equation}\label{eq-lin-vlasov-deriv}
	\begin{split}
	& \left(\frac{\partial}{\partial t} + v_3 \frac{\partial}{\partial x_3}\right) \delta f_s + \frac{\partial}{\partial x_1} \langle v_1 \delta F_{s} \rangle + \frac{\partial}{\partial x_2} \langle v_2 \delta F_{s} \rangle \\
	= & -\frac{q_s}{m_s} \left[\delta E_3 \frac{\partial}{\partial v_3} f_{0s} + \frac{\delta B_2}{c} \frac{\partial}{\partial v_3} \langle v_1 F_{0s} \rangle - \frac{\delta B_1}{c} \frac{\partial}{\partial v_3} \langle v_2 F_{0s} \rangle\right] .
	\end{split}
	\end{equation}
\end{linenomath*}
Here the term with $\partial / \partial v_1$ has been integrated to zero, because the acceleration in front of $\partial / \partial v_1$ is irrelevant to $v_1$ and thus we can directly perform the integration $\int dv_1\, \partial F_{0s} / \partial v_1 = 0$. The same technique has also been applied to the term with $\partial / \partial v_2$. By using the property that $F_{0s}$ is a Maxwellian, we have $\langle v_1 F_{0s} \rangle = 0$ and $\langle v_2 F_{0s} \rangle = 0$. Furthermore, $\delta F_s$ is a function of $\sqrt{v_1^2 + v_2^2}$ but not a function of the gyro-phase. The reason is that the cyclotron resonant velocity is much larger than the electron/ion thermal velocity in this case and virtually no particles are in cyclotron resonance with the kinetic Alfv\'en wave. This leads to $\langle v_1 \delta F_{s} \rangle = 0$ and $\langle v_2 \delta F_{s} \rangle = 0$. With the above considerations, Equation \eqref{eq-lin-vlasov-deriv} can be simplified as
\begin{linenomath*}
	\begin{equation}
		\left(\frac{\partial}{\partial t} + v_3 \frac{\partial}{\partial x_3}\right) \delta f_s = -\frac{q_s \delta E_3}{m_s} \frac{\partial f_{0s}}{\partial v_3} .
	\end{equation}
\end{linenomath*}
This is the linearized Vlasov equation for the reduced particle distribution.

\acknowledgments
This research was supported by NASA Grants NO.~NNX16AG21G and NO.~80NSSC18K1227. The simulation data has been archived on Zenodo \url{https://doi.org/10.5281/zenodo.4005006}. We thank G.~J.~Morales for insightful discussions. We also thank V.~Angelopoulos for explaining the distinction between the nomenclatures ``dipolarization fronts'' and ``injection fronts''. We would like to acknowledge high-performance computing support from Cheyenne (doi:10.5065/D6RX99HX) provided by NCAR's Computational and Information Systems Laboratory, sponsored by the National Science Foundation.  We would also like to acknowledge the \texttt{OSIRIS} Consortium, consisting of UCLA and IST (Lisbon, Portugal) for the use of \texttt{OSIRIS} and for providing access to the \texttt{OSIRIS} 4.0 framework.


%
%

\bibliography{tds}

\begin{thebibliography}{}

\bibitem [\protect \citeauthoryear {%
Agapitov%
\ \protect \BOthers {.}}{%
Agapitov%
\ \protect \BOthers {.}}{%
{\protect \APACyear {2018}}%
}]{%
agapitov2018nonlinear}
\APACinsertmetastar {%
agapitov2018nonlinear}%
\begin{APACrefauthors}%
Agapitov, O.%
, Drake, J.%
, Vasko, I.%
, Mozer, F.%
, Artemyev, A.%
, Krasnoselskikh, V.%
\BDBL {}Reeves, G\BPBI D.%
\end{APACrefauthors}%
\unskip\
\newblock
\APACrefYearMonthDay{2018}{}{}.
\newblock
{\BBOQ}\APACrefatitle {Nonlinear electrostatic steepening of whistler waves:
  The guiding factors and dynamics in inhomogeneous systems} {Nonlinear
  electrostatic steepening of whistler waves: The guiding factors and dynamics
  in inhomogeneous systems}.{\BBCQ}
\newblock
\APACjournalVolNumPages{Geophysical Research Letters}{45}{5}{2168--2176}.
\PrintBackRefs{\CurrentBib}

\bibitem [\protect \citeauthoryear {%
An%
\ \protect \BOthers {.}}{%
An%
\ \protect \BOthers {.}}{%
{\protect \APACyear {2019}}%
}]{%
an2019unified}
\APACinsertmetastar {%
an2019unified}%
\begin{APACrefauthors}%
An, X.%
, Li, J.%
, Bortnik, J.%
, Decyk, V.%
, Kletzing, C.%
\BCBL {}\ \BBA {} Hospodarsky, G.%
\end{APACrefauthors}%
\unskip\
\newblock
\APACrefYearMonthDay{2019}{}{}.
\newblock
{\BBOQ}\APACrefatitle {Unified View of Nonlinear Wave Structures Associated
  with Whistler-Mode Chorus} {Unified view of nonlinear wave structures
  associated with whistler-mode chorus}.{\BBCQ}
\newblock
\APACjournalVolNumPages{Physical review letters}{122}{4}{045101}.
\PrintBackRefs{\CurrentBib}

\bibitem [\protect \citeauthoryear {%
Anderegg%
, Driscoll%
, Dubin%
, O'Neil%
\BCBL {}\ \BBA {} Valentini%
}{%
Anderegg%
\ \protect \BOthers {.}}{%
{\protect \APACyear {2009}}%
}]{%
anderegg2009electron}
\APACinsertmetastar {%
anderegg2009electron}%
\begin{APACrefauthors}%
Anderegg, F.%
, Driscoll, C\BPBI F.%
, Dubin, D\BPBI H.%
, O'Neil, T\BPBI M.%
\BCBL {}\ \BBA {} Valentini, F.%
\end{APACrefauthors}%
\unskip\
\newblock
\APACrefYearMonthDay{2009}{}{}.
\newblock
{\BBOQ}\APACrefatitle {Electron acoustic waves in pure ion plasmas} {Electron
  acoustic waves in pure ion plasmas}.{\BBCQ}
\newblock
\APACjournalVolNumPages{Physics of Plasmas}{16}{5}{055705}.
\PrintBackRefs{\CurrentBib}

\bibitem [\protect \citeauthoryear {%
Artemyev%
, Agapitov%
, Mozer%
\BCBL {}\ \BBA {} Krasnoselskikh%
}{%
Artemyev%
\ \protect \BOthers {.}}{%
{\protect \APACyear {2014}}%
}]{%
artemyev2014thermal}
\APACinsertmetastar {%
artemyev2014thermal}%
\begin{APACrefauthors}%
Artemyev, A.%
, Agapitov, O.%
, Mozer, F.%
\BCBL {}\ \BBA {} Krasnoselskikh, V.%
\end{APACrefauthors}%
\unskip\
\newblock
\APACrefYearMonthDay{2014}{}{}.
\newblock
{\BBOQ}\APACrefatitle {Thermal electron acceleration by localized bursts of
  electric field in the radiation belts} {Thermal electron acceleration by
  localized bursts of electric field in the radiation belts}.{\BBCQ}
\newblock
\APACjournalVolNumPages{Geophysical Research Letters}{41}{16}{5734--5739}.
\PrintBackRefs{\CurrentBib}

\bibitem [\protect \citeauthoryear {%
Artemyev%
, Rankin%
\BCBL {}\ \BBA {} Blanco%
}{%
Artemyev%
\ \protect \BOthers {.}}{%
{\protect \APACyear {2015}}%
}]{%
artemyev2015electron}
\APACinsertmetastar {%
artemyev2015electron}%
\begin{APACrefauthors}%
Artemyev, A.%
, Rankin, R.%
\BCBL {}\ \BBA {} Blanco, M.%
\end{APACrefauthors}%
\unskip\
\newblock
\APACrefYearMonthDay{2015}{}{}.
\newblock
{\BBOQ}\APACrefatitle {Electron trapping and acceleration by kinetic Alfven
  waves in the inner magnetosphere} {Electron trapping and acceleration by
  kinetic alfven waves in the inner magnetosphere}.{\BBCQ}
\newblock
\APACjournalVolNumPages{Journal of Geophysical Research: Space
  Physics}{120}{12}{10--305}.
\PrintBackRefs{\CurrentBib}

\bibitem [\protect \citeauthoryear {%
Artemyev%
, Rankin%
\BCBL {}\ \BBA {} Vasko%
}{%
Artemyev%
\ \protect \BOthers {.}}{%
{\protect \APACyear {2017}}%
}]{%
artemyev2017nonlinear}
\APACinsertmetastar {%
artemyev2017nonlinear}%
\begin{APACrefauthors}%
Artemyev, A.%
, Rankin, R.%
\BCBL {}\ \BBA {} Vasko, I.%
\end{APACrefauthors}%
\unskip\
\newblock
\APACrefYearMonthDay{2017}{}{}.
\newblock
{\BBOQ}\APACrefatitle {Nonlinear Landau resonance with localized wave pulses}
  {Nonlinear landau resonance with localized wave pulses}.{\BBCQ}
\newblock
\APACjournalVolNumPages{Journal of Geophysical Research: Space
  Physics}{122}{5}{5519--5527}.
\PrintBackRefs{\CurrentBib}

\bibitem [\protect \citeauthoryear {%
C.~Chaston%
, Bonnell%
, Clausen%
\BCBL {}\ \BBA {} Angelopoulos%
}{%
C.~Chaston%
\ \protect \BOthers {.}}{%
{\protect \APACyear {2012}}%
}]{%
chaston2012energy}
\APACinsertmetastar {%
chaston2012energy}%
\begin{APACrefauthors}%
Chaston, C.%
, Bonnell, J.%
, Clausen, L.%
\BCBL {}\ \BBA {} Angelopoulos, V.%
\end{APACrefauthors}%
\unskip\
\newblock
\APACrefYearMonthDay{2012}{}{}.
\newblock
{\BBOQ}\APACrefatitle {Energy transport by kinetic-scale electromagnetic waves
  in fast plasma sheet flows} {Energy transport by kinetic-scale
  electromagnetic waves in fast plasma sheet flows}.{\BBCQ}
\newblock
\APACjournalVolNumPages{Journal of Geophysical Research: Space
  Physics}{117}{A9}{}.
\PrintBackRefs{\CurrentBib}

\bibitem [\protect \citeauthoryear {%
C.~Chaston%
\ \protect \BOthers {.}}{%
C.~Chaston%
\ \protect \BOthers {.}}{%
{\protect \APACyear {2015}}%
}]{%
chaston2015broadband}
\APACinsertmetastar {%
chaston2015broadband}%
\begin{APACrefauthors}%
Chaston, C.%
, Bonnell, J.%
, Kletzing, C.%
, Hospodarsky, G.%
, Wygant, J.%
\BCBL {}\ \BBA {} Smith, C.%
\end{APACrefauthors}%
\unskip\
\newblock
\APACrefYearMonthDay{2015}{}{}.
\newblock
{\BBOQ}\APACrefatitle {Broadband low-frequency electromagnetic waves in the
  inner magnetosphere} {Broadband low-frequency electromagnetic waves in the
  inner magnetosphere}.{\BBCQ}
\newblock
\APACjournalVolNumPages{Journal of Geophysical Research: Space
  Physics}{120}{10}{8603--8615}.
\PrintBackRefs{\CurrentBib}

\bibitem [\protect \citeauthoryear {%
C\BPBI C.~Chaston%
\ \protect \BOthers {.}}{%
C\BPBI C.~Chaston%
\ \protect \BOthers {.}}{%
{\protect \APACyear {2014}}%
}]{%
chaston2014observations}
\APACinsertmetastar {%
chaston2014observations}%
\begin{APACrefauthors}%
Chaston, C\BPBI C.%
, Bonnell, J\BPBI W.%
, Wygant, J\BPBI R.%
, Mozer, F.%
, Bale, S\BPBI D.%
, Kersten, K.%
\BDBL {}Macdonald, E\BPBI A.%
\end{APACrefauthors}%
\unskip\
\newblock
\APACrefYearMonthDay{2014}{}{}.
\newblock
{\BBOQ}\APACrefatitle {Observations of kinetic scale field line resonances}
  {Observations of kinetic scale field line resonances}.{\BBCQ}
\newblock
\APACjournalVolNumPages{Geophysical Research Letters}{41}{2}{209--215}.
\PrintBackRefs{\CurrentBib}

\bibitem [\protect \citeauthoryear {%
Damiano%
, Chaston%
, Hull%
\BCBL {}\ \BBA {} Johnson%
}{%
Damiano%
\ \protect \BOthers {.}}{%
{\protect \APACyear {2018}}%
}]{%
damiano2018electron}
\APACinsertmetastar {%
damiano2018electron}%
\begin{APACrefauthors}%
Damiano, P.%
, Chaston, C.%
, Hull, A.%
\BCBL {}\ \BBA {} Johnson, J\BPBI R.%
\end{APACrefauthors}%
\unskip\
\newblock
\APACrefYearMonthDay{2018}{}{}.
\newblock
{\BBOQ}\APACrefatitle {Electron distributions in kinetic scale field line
  resonances: A comparison of simulations and observations} {Electron
  distributions in kinetic scale field line resonances: A comparison of
  simulations and observations}.{\BBCQ}
\newblock
\APACjournalVolNumPages{Geophysical Research Letters}{45}{12}{5826--5835}.
\PrintBackRefs{\CurrentBib}

\bibitem [\protect \citeauthoryear {%
Damiano%
, Johnson%
\BCBL {}\ \BBA {} Chaston%
}{%
Damiano%
\ \protect \BOthers {.}}{%
{\protect \APACyear {2015}}%
}]{%
damiano2015ion}
\APACinsertmetastar {%
damiano2015ion}%
\begin{APACrefauthors}%
Damiano, P.%
, Johnson, J.%
\BCBL {}\ \BBA {} Chaston, C.%
\end{APACrefauthors}%
\unskip\
\newblock
\APACrefYearMonthDay{2015}{}{}.
\newblock
{\BBOQ}\APACrefatitle {Ion temperature effects on magnetotail Alfv{\'e}n wave
  propagation and electron energization} {Ion temperature effects on
  magnetotail alfv{\'e}n wave propagation and electron energization}.{\BBCQ}
\newblock
\APACjournalVolNumPages{Journal of Geophysical Research: Space
  Physics}{120}{7}{5623--5632}.
\PrintBackRefs{\CurrentBib}

\bibitem [\protect \citeauthoryear {%
Damiano%
, Johnson%
\BCBL {}\ \BBA {} Chaston%
}{%
Damiano%
\ \protect \BOthers {.}}{%
{\protect \APACyear {2016}}%
}]{%
damiano2016ion}
\APACinsertmetastar {%
damiano2016ion}%
\begin{APACrefauthors}%
Damiano, P.%
, Johnson, J\BPBI R.%
\BCBL {}\ \BBA {} Chaston, C.%
\end{APACrefauthors}%
\unskip\
\newblock
\APACrefYearMonthDay{2016}{}{}.
\newblock
{\BBOQ}\APACrefatitle {Ion gyroradius effects on particle trapping in kinetic
  Alfv{\'e}n waves along auroral field lines} {Ion gyroradius effects on
  particle trapping in kinetic alfv{\'e}n waves along auroral field
  lines}.{\BBCQ}
\newblock
\APACjournalVolNumPages{Journal of Geophysical Research: Space
  Physics}{121}{11}{10--831}.
\PrintBackRefs{\CurrentBib}

\bibitem [\protect \citeauthoryear {%
Ergun%
, Goodrich%
, Stawarz%
, Andersson%
\BCBL {}\ \BBA {} Angelopoulos%
}{%
Ergun%
\ \protect \BOthers {.}}{%
{\protect \APACyear {2015}}%
}]{%
ergun2015large}
\APACinsertmetastar {%
ergun2015large}%
\begin{APACrefauthors}%
Ergun, R.%
, Goodrich, K.%
, Stawarz, J.%
, Andersson, L.%
\BCBL {}\ \BBA {} Angelopoulos, V.%
\end{APACrefauthors}%
\unskip\
\newblock
\APACrefYearMonthDay{2015}{}{}.
\newblock
{\BBOQ}\APACrefatitle {Large-amplitude electric fields associated with bursty
  bulk flow braking in the Earth's plasma sheet} {Large-amplitude electric
  fields associated with bursty bulk flow braking in the earth's plasma
  sheet}.{\BBCQ}
\newblock
\APACjournalVolNumPages{Journal of Geophysical Research: Space
  Physics}{120}{3}{1832--1844}.
\PrintBackRefs{\CurrentBib}

\bibitem [\protect \citeauthoryear {%
Fonseca%
\ \protect \BOthers {.}}{%
Fonseca%
\ \protect \BOthers {.}}{%
{\protect \APACyear {2002}}%
}]{%
fonseca2002osiris}
\APACinsertmetastar {%
fonseca2002osiris}%
\begin{APACrefauthors}%
Fonseca, R\BPBI A.%
, Silva, L\BPBI O.%
, Tsung, F\BPBI S.%
, Decyk, V\BPBI K.%
, Lu, W.%
, Ren, C.%
\BDBL {}Adam, J\BPBI C.%
\end{APACrefauthors}%
\unskip\
\newblock
\APACrefYearMonthDay{2002}{}{}.
\newblock
{\BBOQ}\APACrefatitle {{OSIRIS}: A three-dimensional, fully relativistic
  particle in cell code for modeling plasma based accelerators} {{OSIRIS}: A
  three-dimensional, fully relativistic particle in cell code for modeling
  plasma based accelerators}.{\BBCQ}
\newblock
\BIn{} \APACrefbtitle {International Conference on Computational Science}
  {International conference on computational science}\ (\BPGS\ 342--351).
\PrintBackRefs{\CurrentBib}

\bibitem [\protect \citeauthoryear {%
Gary%
}{%
Gary%
}{%
{\protect \APACyear {1986}}%
}]{%
gary1986low}
\APACinsertmetastar {%
gary1986low}%
\begin{APACrefauthors}%
Gary, S\BPBI P.%
\end{APACrefauthors}%
\unskip\
\newblock
\APACrefYearMonthDay{1986}{}{}.
\newblock
{\BBOQ}\APACrefatitle {Low-frequency waves in a high-beta collisionless plasma:
  Polarization, compressibility and helicity} {Low-frequency waves in a
  high-beta collisionless plasma: Polarization, compressibility and
  helicity}.{\BBCQ}
\newblock
\APACjournalVolNumPages{Journal of plasma physics}{35}{3}{431--447}.
\PrintBackRefs{\CurrentBib}

\bibitem [\protect \citeauthoryear {%
Gary%
}{%
Gary%
}{%
{\protect \APACyear {1993}}%
}]{%
gary1993theory}
\APACinsertmetastar {%
gary1993theory}%
\begin{APACrefauthors}%
Gary, S\BPBI P.%
\end{APACrefauthors}%
\unskip\
\newblock
\APACrefYear{1993}.
\newblock
\APACrefbtitle {Theory of space plasma microinstabilities} {Theory of space
  plasma microinstabilities}\ (\BNUM~7).
\newblock
\APACaddressPublisher{}{Cambridge university press}.
\PrintBackRefs{\CurrentBib}

\bibitem [\protect \citeauthoryear {%
Gary%
\ \BBA {} Nishimura%
}{%
Gary%
\ \BBA {} Nishimura%
}{%
{\protect \APACyear {2004}}%
}]{%
gary2004kinetic}
\APACinsertmetastar {%
gary2004kinetic}%
\begin{APACrefauthors}%
Gary, S\BPBI P.%
\BCBT {}\ \BBA {} Nishimura, K.%
\end{APACrefauthors}%
\unskip\
\newblock
\APACrefYearMonthDay{2004}{}{}.
\newblock
{\BBOQ}\APACrefatitle {Kinetic Alfv{\'e}n waves: Linear theory and a
  particle-in-cell simulation} {Kinetic alfv{\'e}n waves: Linear theory and a
  particle-in-cell simulation}.{\BBCQ}
\newblock
\APACjournalVolNumPages{Journal of Geophysical Research: Space
  Physics}{109}{A2}{}.
\PrintBackRefs{\CurrentBib}

\bibitem [\protect \citeauthoryear {%
G{\'e}not%
, Louarn%
\BCBL {}\ \BBA {} Mottez%
}{%
G{\'e}not%
\ \protect \BOthers {.}}{%
{\protect \APACyear {2004}}%
}]{%
genot2004alfven}
\APACinsertmetastar {%
genot2004alfven}%
\begin{APACrefauthors}%
G{\'e}not, V.%
, Louarn, P.%
\BCBL {}\ \BBA {} Mottez, F.%
\end{APACrefauthors}%
\unskip\
\newblock
\APACrefYearMonthDay{2004}{}{}.
\newblock
{\BBOQ}\APACrefatitle {Alfv{\'e}n wave interaction with inhomogeneous plasmas:
  acceleration and energy cascade towards small-scales} {Alfv{\'e}n wave
  interaction with inhomogeneous plasmas: acceleration and energy cascade
  towards small-scales}.{\BBCQ}
\newblock
\BIn{} \APACrefbtitle {Annales Geophysicae} {Annales geophysicae}\ (\BVOL~22,
  \BPGS\ 2081--2096).
\PrintBackRefs{\CurrentBib}

\bibitem [\protect \citeauthoryear {%
Goertz%
\ \BBA {} Boswell%
}{%
Goertz%
\ \BBA {} Boswell%
}{%
{\protect \APACyear {1979}}%
}]{%
goertz1979magnetosphere}
\APACinsertmetastar {%
goertz1979magnetosphere}%
\begin{APACrefauthors}%
Goertz, C.%
\BCBT {}\ \BBA {} Boswell, R.%
\end{APACrefauthors}%
\unskip\
\newblock
\APACrefYearMonthDay{1979}{}{}.
\newblock
{\BBOQ}\APACrefatitle {Magnetosphere-ionosphere coupling}
  {Magnetosphere-ionosphere coupling}.{\BBCQ}
\newblock
\APACjournalVolNumPages{Journal of Geophysical Research: Space
  Physics}{84}{A12}{7239--7246}.
\PrintBackRefs{\CurrentBib}

\bibitem [\protect \citeauthoryear {%
Hasegawa%
}{%
Hasegawa%
}{%
{\protect \APACyear {1976}}%
}]{%
hasegawa1976particle}
\APACinsertmetastar {%
hasegawa1976particle}%
\begin{APACrefauthors}%
Hasegawa, A.%
\end{APACrefauthors}%
\unskip\
\newblock
\APACrefYearMonthDay{1976}{}{}.
\newblock
{\BBOQ}\APACrefatitle {Particle acceleration by MHD surface wave and formation
  of aurora} {Particle acceleration by mhd surface wave and formation of
  aurora}.{\BBCQ}
\newblock
\APACjournalVolNumPages{Journal of Geophysical Research}{81}{28}{5083--5090}.
\PrintBackRefs{\CurrentBib}

\bibitem [\protect \citeauthoryear {%
Hemker%
}{%
Hemker%
}{%
{\protect \APACyear {2015}}%
}]{%
hemker2015particle}
\APACinsertmetastar {%
hemker2015particle}%
\begin{APACrefauthors}%
Hemker, R\BPBI G.%
\end{APACrefauthors}%
\unskip\
\newblock
\APACrefYearMonthDay{2015}{}{}.
\newblock
{\BBOQ}\APACrefatitle {Particle-in-cell modeling of plasma-based accelerators
  in two and three dimensions} {Particle-in-cell modeling of plasma-based
  accelerators in two and three dimensions}.{\BBCQ}
\newblock
\APACjournalVolNumPages{arXiv preprint arXiv:1503.00276}{}{}{}.
\PrintBackRefs{\CurrentBib}

\bibitem [\protect \citeauthoryear {%
Holloway%
\ \BBA {} Dorning%
}{%
Holloway%
\ \BBA {} Dorning%
}{%
{\protect \APACyear {1991}}%
}]{%
holloway1991undamped}
\APACinsertmetastar {%
holloway1991undamped}%
\begin{APACrefauthors}%
Holloway, J\BPBI P.%
\BCBT {}\ \BBA {} Dorning, J.%
\end{APACrefauthors}%
\unskip\
\newblock
\APACrefYearMonthDay{1991}{}{}.
\newblock
{\BBOQ}\APACrefatitle {Undamped plasma waves} {Undamped plasma waves}.{\BBCQ}
\newblock
\APACjournalVolNumPages{Physical Review A}{44}{6}{3856}.
\PrintBackRefs{\CurrentBib}

\bibitem [\protect \citeauthoryear {%
Horne%
}{%
Horne%
}{%
{\protect \APACyear {1989}}%
}]{%
horne1989path}
\APACinsertmetastar {%
horne1989path}%
\begin{APACrefauthors}%
Horne, R\BPBI B.%
\end{APACrefauthors}%
\unskip\
\newblock
\APACrefYearMonthDay{1989}{}{}.
\newblock
{\BBOQ}\APACrefatitle {Path-integrated growth of electrostatic waves: The
  generation of terrestrial myriametric radiation} {Path-integrated growth of
  electrostatic waves: The generation of terrestrial myriametric
  radiation}.{\BBCQ}
\newblock
\APACjournalVolNumPages{Journal of Geophysical Research: Space
  Physics}{94}{A7}{8895--8909}.
\PrintBackRefs{\CurrentBib}

\bibitem [\protect \citeauthoryear {%
Ichimaru%
}{%
Ichimaru%
}{%
{\protect \APACyear {2018}}%
}]{%
ichimaru2018basic}
\APACinsertmetastar {%
ichimaru2018basic}%
\begin{APACrefauthors}%
Ichimaru, S.%
\end{APACrefauthors}%
\unskip\
\newblock
\APACrefYear{2018}.
\newblock
\APACrefbtitle {Basic principles of plasma physics: a statistical approach}
  {Basic principles of plasma physics: a statistical approach}.
\newblock
\APACaddressPublisher{}{CRC Press}.
\PrintBackRefs{\CurrentBib}

\bibitem [\protect \citeauthoryear {%
Li%
\ \protect \BOthers {.}}{%
Li%
\ \protect \BOthers {.}}{%
{\protect \APACyear {2017}}%
}]{%
li2017chorus}
\APACinsertmetastar {%
li2017chorus}%
\begin{APACrefauthors}%
Li, J.%
, Bortnik, J.%
, An, X.%
, Li, W.%
, Thorne, R\BPBI M.%
, Zhou, M.%
\BDBL {}Spence, H\BPBI E.%
\end{APACrefauthors}%
\unskip\
\newblock
\APACrefYearMonthDay{2017}{}{}.
\newblock
{\BBOQ}\APACrefatitle {Chorus wave modulation of Langmuir waves in the
  radiation belts} {Chorus wave modulation of langmuir waves in the radiation
  belts}.{\BBCQ}
\newblock
\APACjournalVolNumPages{Geophysical Research Letters}{44}{23}{11--713}.
\PrintBackRefs{\CurrentBib}

\bibitem [\protect \citeauthoryear {%
Lysak%
\ \BBA {} Lotko%
}{%
Lysak%
\ \BBA {} Lotko%
}{%
{\protect \APACyear {1996}}%
}]{%
lysak1996kinetic}
\APACinsertmetastar {%
lysak1996kinetic}%
\begin{APACrefauthors}%
Lysak, R\BPBI L.%
\BCBT {}\ \BBA {} Lotko, W.%
\end{APACrefauthors}%
\unskip\
\newblock
\APACrefYearMonthDay{1996}{}{}.
\newblock
{\BBOQ}\APACrefatitle {On the kinetic dispersion relation for shear Alfv{\'e}n
  waves} {On the kinetic dispersion relation for shear alfv{\'e}n
  waves}.{\BBCQ}
\newblock
\APACjournalVolNumPages{Journal of Geophysical Research: Space
  Physics}{101}{A3}{5085--5094}.
\PrintBackRefs{\CurrentBib}

\bibitem [\protect \citeauthoryear {%
Malaspina%
\ \protect \BOthers {.}}{%
Malaspina%
\ \protect \BOthers {.}}{%
{\protect \APACyear {2014}}%
}]{%
malaspina2014nonlinear}
\APACinsertmetastar {%
malaspina2014nonlinear}%
\begin{APACrefauthors}%
Malaspina, D\BPBI M.%
, Andersson, L.%
, Ergun, R\BPBI E.%
, Wygant, J\BPBI R.%
, Bonnell, J.%
, Kletzing, C.%
\BDBL {}Larsen, B\BPBI A.%
\end{APACrefauthors}%
\unskip\
\newblock
\APACrefYearMonthDay{2014}{}{}.
\newblock
{\BBOQ}\APACrefatitle {Nonlinear electric field structures in the inner
  magnetosphere} {Nonlinear electric field structures in the inner
  magnetosphere}.{\BBCQ}
\newblock
\APACjournalVolNumPages{Geophysical Research Letters}{41}{16}{5693--5701}.
\PrintBackRefs{\CurrentBib}

\bibitem [\protect \citeauthoryear {%
Malaspina%
, Ukhorskiy%
, Chu%
\BCBL {}\ \BBA {} Wygant%
}{%
Malaspina%
\ \protect \BOthers {.}}{%
{\protect \APACyear {2018}}%
}]{%
malaspina2018census}
\APACinsertmetastar {%
malaspina2018census}%
\begin{APACrefauthors}%
Malaspina, D\BPBI M.%
, Ukhorskiy, A.%
, Chu, X.%
\BCBL {}\ \BBA {} Wygant, J.%
\end{APACrefauthors}%
\unskip\
\newblock
\APACrefYearMonthDay{2018}{}{}.
\newblock
{\BBOQ}\APACrefatitle {A census of plasma waves and structures associated with
  an injection front in the inner magnetosphere} {A census of plasma waves and
  structures associated with an injection front in the inner
  magnetosphere}.{\BBCQ}
\newblock
\APACjournalVolNumPages{Journal of Geophysical Research: Space
  Physics}{123}{4}{2566--2587}.
\PrintBackRefs{\CurrentBib}

\bibitem [\protect \citeauthoryear {%
Malaspina%
\ \protect \BOthers {.}}{%
Malaspina%
\ \protect \BOthers {.}}{%
{\protect \APACyear {2015}}%
}]{%
malaspina2015electric}
\APACinsertmetastar {%
malaspina2015electric}%
\begin{APACrefauthors}%
Malaspina, D\BPBI M.%
, Wygant, J\BPBI R.%
, Ergun, R\BPBI E.%
, Reeves, G\BPBI D.%
, Skoug, R\BPBI M.%
\BCBL {}\ \BBA {} Larsen, B\BPBI A.%
\end{APACrefauthors}%
\unskip\
\newblock
\APACrefYearMonthDay{2015}{}{}.
\newblock
{\BBOQ}\APACrefatitle {Electric field structures and waves at plasma boundaries
  in the inner magnetosphere} {Electric field structures and waves at plasma
  boundaries in the inner magnetosphere}.{\BBCQ}
\newblock
\APACjournalVolNumPages{Journal of Geophysical Research: Space
  Physics}{120}{6}{4246--4263}.
\PrintBackRefs{\CurrentBib}

\bibitem [\protect \citeauthoryear {%
Mozer%
\ \protect \BOthers {.}}{%
Mozer%
\ \protect \BOthers {.}}{%
{\protect \APACyear {2015}}%
}]{%
mozer2015time}
\APACinsertmetastar {%
mozer2015time}%
\begin{APACrefauthors}%
Mozer, F.%
, Agapitov, O.%
, Artemyev, A.%
, Drake, J.%
, Krasnoselskikh, V.%
, Lejosne, S.%
\BCBL {}\ \BBA {} Vasko, I.%
\end{APACrefauthors}%
\unskip\
\newblock
\APACrefYearMonthDay{2015}{}{}.
\newblock
{\BBOQ}\APACrefatitle {Time domain structures: What and where they are, what
  they do, and how they are made} {Time domain structures: What and where they
  are, what they do, and how they are made}.{\BBCQ}
\newblock
\APACjournalVolNumPages{Geophysical Research Letters}{42}{10}{3627--3638}.
\PrintBackRefs{\CurrentBib}

\bibitem [\protect \citeauthoryear {%
O'Neil%
}{%
O'Neil%
}{%
{\protect \APACyear {1965}}%
}]{%
o1965collisionless}
\APACinsertmetastar {%
o1965collisionless}%
\begin{APACrefauthors}%
O'Neil, T.%
\end{APACrefauthors}%
\unskip\
\newblock
\APACrefYearMonthDay{1965}{}{}.
\newblock
{\BBOQ}\APACrefatitle {Collisionless damping of nonlinear plasma oscillations}
  {Collisionless damping of nonlinear plasma oscillations}.{\BBCQ}
\newblock
\APACjournalVolNumPages{The physics of fluids}{8}{12}{2255--2262}.
\PrintBackRefs{\CurrentBib}

\bibitem [\protect \citeauthoryear {%
O'Neil%
\ \BBA {} Malmberg%
}{%
O'Neil%
\ \BBA {} Malmberg%
}{%
{\protect \APACyear {1968}}%
}]{%
o1968transition}
\APACinsertmetastar {%
o1968transition}%
\begin{APACrefauthors}%
O'Neil, T.%
\BCBT {}\ \BBA {} Malmberg, J.%
\end{APACrefauthors}%
\unskip\
\newblock
\APACrefYearMonthDay{1968}{}{}.
\newblock
{\BBOQ}\APACrefatitle {Transition of the Dispersion Roots from Beam-Type to
  Landau-Type Solutions} {Transition of the dispersion roots from beam-type to
  landau-type solutions}.{\BBCQ}
\newblock
\APACjournalVolNumPages{The Physics of Fluids}{11}{8}{1754--1760}.
\PrintBackRefs{\CurrentBib}

\bibitem [\protect \citeauthoryear {%
Osmane%
\ \BBA {} Pulkkinen%
}{%
Osmane%
\ \BBA {} Pulkkinen%
}{%
{\protect \APACyear {2014}}%
}]{%
osmane2014threshold}
\APACinsertmetastar {%
osmane2014threshold}%
\begin{APACrefauthors}%
Osmane, A.%
\BCBT {}\ \BBA {} Pulkkinen, T\BPBI I.%
\end{APACrefauthors}%
\unskip\
\newblock
\APACrefYearMonthDay{2014}{}{}.
\newblock
{\BBOQ}\APACrefatitle {On the threshold energization of radiation belt
  electrons by double layers} {On the threshold energization of radiation belt
  electrons by double layers}.{\BBCQ}
\newblock
\APACjournalVolNumPages{Journal of Geophysical Research: Space
  Physics}{119}{10}{8243--8248}.
\PrintBackRefs{\CurrentBib}

\bibitem [\protect \citeauthoryear {%
Quon%
\ \BBA {} Wong%
}{%
Quon%
\ \BBA {} Wong%
}{%
{\protect \APACyear {1976}}%
}]{%
quon1976formation}
\APACinsertmetastar {%
quon1976formation}%
\begin{APACrefauthors}%
Quon, B.%
\BCBT {}\ \BBA {} Wong, A.%
\end{APACrefauthors}%
\unskip\
\newblock
\APACrefYearMonthDay{1976}{}{}.
\newblock
{\BBOQ}\APACrefatitle {Formation of potential double layers in plasmas}
  {Formation of potential double layers in plasmas}.{\BBCQ}
\newblock
\APACjournalVolNumPages{Physical Review Letters}{37}{21}{1393}.
\PrintBackRefs{\CurrentBib}

\bibitem [\protect \citeauthoryear {%
Reinleitner%
, Gurnett%
\BCBL {}\ \BBA {} Gallagher%
}{%
Reinleitner%
\ \protect \BOthers {.}}{%
{\protect \APACyear {1982}}%
}]{%
reinleitner1982chorus}
\APACinsertmetastar {%
reinleitner1982chorus}%
\begin{APACrefauthors}%
Reinleitner, L\BPBI A.%
, Gurnett, D\BPBI A.%
\BCBL {}\ \BBA {} Gallagher, D\BPBI L.%
\end{APACrefauthors}%
\unskip\
\newblock
\APACrefYearMonthDay{1982}{}{}.
\newblock
{\BBOQ}\APACrefatitle {Chorus-related electrostatic bursts in the {E}arth's
  outer magnetosphere} {Chorus-related electrostatic bursts in the {E}arth's
  outer magnetosphere}.{\BBCQ}
\newblock
\APACjournalVolNumPages{Nature}{295}{5844}{46}.
\PrintBackRefs{\CurrentBib}

\bibitem [\protect \citeauthoryear {%
Schamel%
}{%
Schamel%
}{%
{\protect \APACyear {1979}}%
}]{%
schamel1979theory}
\APACinsertmetastar {%
schamel1979theory}%
\begin{APACrefauthors}%
Schamel, H.%
\end{APACrefauthors}%
\unskip\
\newblock
\APACrefYearMonthDay{1979}{}{}.
\newblock
{\BBOQ}\APACrefatitle {Theory of electron holes} {Theory of electron
  holes}.{\BBCQ}
\newblock
\APACjournalVolNumPages{Physica Scripta}{20}{3-4}{336}.
\PrintBackRefs{\CurrentBib}

\bibitem [\protect \citeauthoryear {%
Silberstein%
\ \BBA {} Otani%
}{%
Silberstein%
\ \BBA {} Otani%
}{%
{\protect \APACyear {1994}}%
}]{%
silberstein1994computer}
\APACinsertmetastar {%
silberstein1994computer}%
\begin{APACrefauthors}%
Silberstein, M.%
\BCBT {}\ \BBA {} Otani, N.%
\end{APACrefauthors}%
\unskip\
\newblock
\APACrefYearMonthDay{1994}{}{}.
\newblock
{\BBOQ}\APACrefatitle {Computer simulation of Alfv{\'e}n waves and double
  layers along auroral magnetic field lines} {Computer simulation of alfv{\'e}n
  waves and double layers along auroral magnetic field lines}.{\BBCQ}
\newblock
\APACjournalVolNumPages{Journal of Geophysical Research: Space
  Physics}{99}{A4}{6351--6365}.
\PrintBackRefs{\CurrentBib}

\bibitem [\protect \citeauthoryear {%
Stawarz%
, Ergun%
\BCBL {}\ \BBA {} Goodrich%
}{%
Stawarz%
\ \protect \BOthers {.}}{%
{\protect \APACyear {2015}}%
}]{%
stawarz2015generation}
\APACinsertmetastar {%
stawarz2015generation}%
\begin{APACrefauthors}%
Stawarz, J.%
, Ergun, R.%
\BCBL {}\ \BBA {} Goodrich, K.%
\end{APACrefauthors}%
\unskip\
\newblock
\APACrefYearMonthDay{2015}{}{}.
\newblock
{\BBOQ}\APACrefatitle {Generation of high-frequency electric field activity by
  turbulence in the Earth's magnetotail} {Generation of high-frequency electric
  field activity by turbulence in the earth's magnetotail}.{\BBCQ}
\newblock
\APACjournalVolNumPages{Journal of Geophysical Research: Space
  Physics}{120}{3}{1845--1866}.
\PrintBackRefs{\CurrentBib}

\bibitem [\protect \citeauthoryear {%
Stix%
}{%
Stix%
}{%
{\protect \APACyear {1992}}%
}]{%
stix1992waves}
\APACinsertmetastar {%
stix1992waves}%
\begin{APACrefauthors}%
Stix, T\BPBI H.%
\end{APACrefauthors}%
\unskip\
\newblock
\APACrefYear{1992}.
\newblock
\APACrefbtitle {Waves in plasmas} {Waves in plasmas}.
\newblock
\APACaddressPublisher{}{Springer Science \& Business Media}.
\PrintBackRefs{\CurrentBib}

\bibitem [\protect \citeauthoryear {%
Swanson%
}{%
Swanson%
}{%
{\protect \APACyear {2012}}%
}]{%
swanson2012plasma}
\APACinsertmetastar {%
swanson2012plasma}%
\begin{APACrefauthors}%
Swanson, D\BPBI G.%
\end{APACrefauthors}%
\unskip\
\newblock
\APACrefYear{2012}.
\newblock
\APACrefbtitle {Plasma waves} {Plasma waves}.
\newblock
\APACaddressPublisher{}{Elsevier}.
\PrintBackRefs{\CurrentBib}

\bibitem [\protect \citeauthoryear {%
A.~Ukhorskiy%
\ \protect \BOthers {.}}{%
A.~Ukhorskiy%
\ \protect \BOthers {.}}{%
{\protect \APACyear {2019}}%
}]{%
ukhorskiy2019kinetic}
\APACinsertmetastar {%
ukhorskiy2019kinetic}%
\begin{APACrefauthors}%
Ukhorskiy, A.%
, Sorathia, K.%
, Merkin, V.%
, Crabtree, C.%
, Fletcher, A.%
\BCBL {}\ \BBA {} Malaspina, D.%
\end{APACrefauthors}%
\unskip\
\newblock
\APACrefYearMonthDay{2019}{}{}.
\newblock
{\BBOQ}\APACrefatitle {Kinetic Properties of Mesoscale Plasma Injections}
  {Kinetic properties of mesoscale plasma injections}.{\BBCQ}
\newblock
\BIn{} \APACrefbtitle {2019 International Conference on Electromagnetics in
  Advanced Applications (ICEAA)} {2019 international conference on
  electromagnetics in advanced applications (iceaa)}\ (\BPGS\ 1350--1350).
\PrintBackRefs{\CurrentBib}

\bibitem [\protect \citeauthoryear {%
A\BPBI Y.~Ukhorskiy%
\ \protect \BOthers {.}}{%
A\BPBI Y.~Ukhorskiy%
\ \protect \BOthers {.}}{%
{\protect \APACyear {2018}}%
}]{%
ukhorskiy2018microscopic}
\APACinsertmetastar {%
ukhorskiy2018microscopic}%
\begin{APACrefauthors}%
Ukhorskiy, A\BPBI Y.%
, Sorathia, K.%
, Merkin, V\BPBI G.%
, Crabtree, C\BPBI E.%
, Fletcher, A.%
\BCBL {}\ \BBA {} Malaspina, D.%
\end{APACrefauthors}%
\unskip\
\newblock
\APACrefYearMonthDay{2018}{}{}.
\newblock
{\BBOQ}\APACrefatitle {Microscopic Properties of Mesoscale Plasma Injections}
  {Microscopic properties of mesoscale plasma injections}.{\BBCQ}
\newblock
\BIn{} \APACrefbtitle {AGU Fall Meeting 2018.} {Agu fall meeting 2018.}
\PrintBackRefs{\CurrentBib}

\bibitem [\protect \citeauthoryear {%
Valentini%
, O'Neil%
\BCBL {}\ \BBA {} Dubin%
}{%
Valentini%
\ \protect \BOthers {.}}{%
{\protect \APACyear {2006}}%
}]{%
valentini2006excitation}
\APACinsertmetastar {%
valentini2006excitation}%
\begin{APACrefauthors}%
Valentini, F.%
, O'Neil, T\BPBI M.%
\BCBL {}\ \BBA {} Dubin, D\BPBI H.%
\end{APACrefauthors}%
\unskip\
\newblock
\APACrefYearMonthDay{2006}{}{}.
\newblock
{\BBOQ}\APACrefatitle {Excitation of nonlinear electron acoustic waves}
  {Excitation of nonlinear electron acoustic waves}.{\BBCQ}
\newblock
\APACjournalVolNumPages{Physics of plasmas}{13}{5}{052303}.
\PrintBackRefs{\CurrentBib}

\bibitem [\protect \citeauthoryear {%
I.~Vasko%
, Agapitov%
, Mozer%
, Artemyev%
, Drake%
\BCBL {}\ \BBA {} Kuzichev%
}{%
I.~Vasko%
, Agapitov%
, Mozer%
, Artemyev%
, Drake%
\BCBL {}\ \BBA {} Kuzichev%
}{%
{\protect \APACyear {2017}}%
}]{%
vasko2017electron}
\APACinsertmetastar {%
vasko2017electron}%
\begin{APACrefauthors}%
Vasko, I.%
, Agapitov, O.%
, Mozer, F.%
, Artemyev, A.%
, Drake, J.%
\BCBL {}\ \BBA {} Kuzichev, I.%
\end{APACrefauthors}%
\unskip\
\newblock
\APACrefYearMonthDay{2017}{}{}.
\newblock
{\BBOQ}\APACrefatitle {Electron holes in the outer radiation belt:
  Characteristics and their role in electron energization} {Electron holes in
  the outer radiation belt: Characteristics and their role in electron
  energization}.{\BBCQ}
\newblock
\APACjournalVolNumPages{Journal of Geophysical Research: Space
  Physics}{122}{1}{120--135}.
\PrintBackRefs{\CurrentBib}

\bibitem [\protect \citeauthoryear {%
I.~Vasko%
, Agapitov%
, Mozer%
, Artemyev%
, Krasnoselskikh%
\BCBL {}\ \BBA {} Bonnell%
}{%
I.~Vasko%
, Agapitov%
, Mozer%
, Artemyev%
, Krasnoselskikh%
\BCBL {}\ \BBA {} Bonnell%
}{%
{\protect \APACyear {2017}}%
}]{%
vasko2017diffusive}
\APACinsertmetastar {%
vasko2017diffusive}%
\begin{APACrefauthors}%
Vasko, I.%
, Agapitov, O.%
, Mozer, F.%
, Artemyev, A.%
, Krasnoselskikh, V.%
\BCBL {}\ \BBA {} Bonnell, J.%
\end{APACrefauthors}%
\unskip\
\newblock
\APACrefYearMonthDay{2017}{}{}.
\newblock
{\BBOQ}\APACrefatitle {Diffusive scattering of electrons by electron holes
  around injection fronts} {Diffusive scattering of electrons by electron holes
  around injection fronts}.{\BBCQ}
\newblock
\APACjournalVolNumPages{Journal of Geophysical Research: Space
  Physics}{122}{3}{3163--3182}.
\PrintBackRefs{\CurrentBib}

\bibitem [\protect \citeauthoryear {%
I.~Vasko%
, Agapitov%
, Mozer%
, Bonnell%
\BCBL {}\ \protect \BOthers {.}}{%
I.~Vasko%
, Agapitov%
, Mozer%
, Bonnell%
\BCBL {}\ \protect \BOthers {.}}{%
{\protect \APACyear {2017}}%
}]{%
vasko2017electron2}
\APACinsertmetastar {%
vasko2017electron2}%
\begin{APACrefauthors}%
Vasko, I.%
, Agapitov, O.%
, Mozer, F.%
, Bonnell, J.%
, Artemyev, A.%
, Krasnoselskikh, V.%
\BDBL {}Hospodarsky, G.%
\end{APACrefauthors}%
\unskip\
\newblock
\APACrefYearMonthDay{2017}{}{}.
\newblock
{\BBOQ}\APACrefatitle {Electron-acoustic solitons and double layers in the
  inner magnetosphere} {Electron-acoustic solitons and double layers in the
  inner magnetosphere}.{\BBCQ}
\newblock
\APACjournalVolNumPages{Geophysical Research Letters}{44}{10}{4575--4583}.
\PrintBackRefs{\CurrentBib}

\bibitem [\protect \citeauthoryear {%
I\BPBI Y.~Vasko%
\ \protect \BOthers {.}}{%
I\BPBI Y.~Vasko%
\ \protect \BOthers {.}}{%
{\protect \APACyear {2018}}%
}]{%
vasko2018electrostatic}
\APACinsertmetastar {%
vasko2018electrostatic}%
\begin{APACrefauthors}%
Vasko, I\BPBI Y.%
, Agapitov, O\BPBI V.%
, Mozer, F\BPBI S.%
, Bonnell, J\BPBI W.%
, Artemyev, A\BPBI V.%
, Krasnoselskikh, V\BPBI V.%
\BCBL {}\ \BBA {} Tong, Y.%
\end{APACrefauthors}%
\unskip\
\newblock
\APACrefYearMonthDay{2018}{}{}.
\newblock
{\BBOQ}\APACrefatitle {Electrostatic steepening of whistler waves}
  {Electrostatic steepening of whistler waves}.{\BBCQ}
\newblock
\APACjournalVolNumPages{Physical review letters}{120}{19}{195101}.
\PrintBackRefs{\CurrentBib}

\bibitem [\protect \citeauthoryear {%
Walsh%
, Hull%
, Agapitov%
, Mozer%
\BCBL {}\ \BBA {} Li%
}{%
Walsh%
\ \protect \BOthers {.}}{%
{\protect \APACyear {2020}}%
}]{%
walsh2020census}
\APACinsertmetastar {%
walsh2020census}%
\begin{APACrefauthors}%
Walsh, B\BPBI M.%
, Hull, A.%
, Agapitov, O.%
, Mozer, F\BPBI S.%
\BCBL {}\ \BBA {} Li, H.%
\end{APACrefauthors}%
\unskip\
\newblock
\APACrefYearMonthDay{2020}{}{}.
\newblock
{\BBOQ}\APACrefatitle {A census of magnetospheric electrons from several eV to
  30 keV} {A census of magnetospheric electrons from several ev to 30
  kev}.{\BBCQ}
\newblock
\APACjournalVolNumPages{Journal of Geophysical Research: Space
  Physics}{}{}{e2019JA027577}.
\PrintBackRefs{\CurrentBib}

\bibitem [\protect \citeauthoryear {%
Weber%
\ \BBA {} Arfken%
}{%
Weber%
\ \BBA {} Arfken%
}{%
{\protect \APACyear {2003}}%
}]{%
weber2003essential}
\APACinsertmetastar {%
weber2003essential}%
\begin{APACrefauthors}%
Weber, H\BPBI J.%
\BCBT {}\ \BBA {} Arfken, G\BPBI B.%
\end{APACrefauthors}%
\unskip\
\newblock
\APACrefYear{2003}.
\newblock
\APACrefbtitle {Essential mathematical methods for physicists, ISE} {Essential
  mathematical methods for physicists, ise}.
\newblock
\APACaddressPublisher{}{Elsevier}.
\PrintBackRefs{\CurrentBib}

\end{thebibliography}

%
%
%
%
%

\end{document}